\documentclass[prc,aps,floats,floatfix,twocolumn,showpacs,nofootinbib,%
superscriptaddress]{revtex4}

\usepackage{amsmath}
\usepackage{amssymb}
\usepackage{amsfonts}
\usepackage{graphicx}
\usepackage{tabularx}
\usepackage{multirow}
\usepackage{lscape}
\usepackage{rotating}

\usepackage{pstricks}

\graphicspath{{plots/}}

\begin{document}

\title{Linear response theory and neutrino mean free path using Brussels-Montreal Skyrme functionals}


\author{A. Pastore}
\affiliation{Institut d'Astronomie et d'Astrophysique, Code Postal 226, Universit\'e Libre de Bruxelles, B-1050 Brussels, Belgium}
\author{M. Martini}
\affiliation{Department of Physics and Astronomy, Ghent University, Proeftuinstraat 86, B-9000 Gent, Belgium}
\affiliation{Institut d'Astronomie et d'Astrophysique, Code Postal 226, Universit\'e Libre de Bruxelles, B-1050 Brussels, Belgium}

\author{D. Davesne}
\affiliation{Universit\'e de Lyon, F-69003 Lyon, France; Universit\'e Lyon 1,
             43 Bd. du 11 Novembre 1918, F-69622 Villeurbanne Cedex, France   \\
             CNRS-IN2P3, UMR 5822, Institut de Physique Nucl{\'e}aire de Lyon}
             
\author{J. Navarro}
\affiliation{IFIC (CSIC-Universidad de Valencia), Apdo. 22085, E-46.071-Valencia, Spain}
             
\author{S. Goriely}
\affiliation{Institut d'Astronomie et d'Astrophysique, Code Postal 226, Universit\'e Libre de Bruxelles, B-1050 Brussels, Belgium}

\author{N. Chamel}
\affiliation{Institut d'Astronomie et d'Astrophysique, Code Postal 226, Universit\'e Libre de Bruxelles, B-1050 Brussels, Belgium}



\begin{abstract}
The Brussels-Montreal Skyrme functionals have been successful to describe properties of both finite nuclei and infinite homogeneous nuclear matter. In their latest version, these functionals have been equipped with two extra density-dependent terms 
in order to reproduce simultaneously ground state properties of nuclei and infinite nuclear matter properties while avoiding at the same time the arising of ferromagnetic instabilities.
In the present article, we extend our previous results of the linear response theory to include such extra terms at both zero and finite temperature in pure neutron matter. The resulting formalism is then applied to derive the neutrino mean free path. The predictions from the Brussels-Montreal Skyrme functionals  are compared with {\it ab-initio} methods.
\end{abstract}


\pacs{21.30.Fe, 21.60.Jz, 21.65.Cd, 26.60.Kp}
 
\date{\today}


\maketitle

%
\section{Introduction}
\label{sect:intro}

The nuclear energy density functional (NEDF) derived from Skyrme effective interactions has been very successful in describing the structure and the dynamics of medium-mass and heavy nuclei~\cite{ben03}.
Thanks to its $universal$ character, the NEDF has also been widely used to study the properties of neutron stars~\cite{dut12}. In particular,
the construction of a universal Skyrme functional which could be applied for the description of various nuclear systems, from finite nuclei to neutron stars,
has been the motivation at the basis of the Brussels-Montreal, BSk, forces~\cite{pea13B}.
These functionals~\cite{gor09,cha09,gor10} have not only been fitted to ground-state properties of $\approx$2000 measured nuclei (masses and radii)~\cite{audi03}, but have also been constrained to reproduce properties of infinite nuclear matter, as obtained from realistic many-body calculations.
The properties of Skyrme functionals in infinite matter have been  investigated showing the appearence of a ferromagnetic instability~\cite{vid84,mar02,cha10B}. It is nowadays agreed that this ferromagnetic instability is a general pathology of the Skyrme functional which is not found in {\it ab-initio} calculations~\cite{fan01,agu14}.
To better satisfy all these constraints and to avoid this spurious phase transition, the BSk family has been recently equipped with some extra density-dependent terms~\cite{cha09,gor10}. The resulting functionals have been employed to compute the composition and the equation of state of dense matter in all regions of a neutron star~\cite{fan13}.

An interesting application of the NEDF theory is the calculation of transport properties of neutrino  in both supernovae and neutron stars~\cite{iwa82}.
In a recent article, we have studied the properties of neutrino mean free path (NMFP) in cold neutron matter~\cite{pas12a} using the $TIJ$-Skyrme family~\cite{les07}. 
The neutrino cross sections in nuclear matter play a central role in determining neutrino transport properties.
During the gravitational collapse of a massive star, energy accumulates in the core.
A very efficient process to evacuate this energy and reach a thermal equilibrium is constituted by the emission of neutrinos.
We have to recall that these neutrinos have to cross several layers of nuclear matter, where a neutrino $\nu$ is scattered by a nucleon $n$ as
\begin{equation}
n+\nu \longrightarrow n'+\nu',
\end{equation}
\noindent   and thus  looses energy. As a consequence, this process reduces the efficiency of energy dissipation through neutrino emission.
The determination of this scattering cross section is  important for astrophysical models~\cite{red98} and  has been  calculated using different nuclear models, such  as Skyrme~\cite{nav99,red98,mar06,pas12a} or \emph{ab-initio} methods ~\cite{cow04,shen03,bac09,lov14} among others, but also relativistic calculations~\cite{red98,hut04,nie01,gry10}.

In the present article, we study the properties of hot neutron matter  when excited by an external probe, by means of the linear response (LR) formalism with the recent BSk functionals. The explicit calculation of the response function of the system will allow us to determine the mean free path of low energetic neutrinos crossing a layer of pure neutron matter (PNM). 


The article is organized as follows: in Sec.\ref{sec:form} we discuss the extra terms introduced into the Skyrme functional and the main ground state properties of the infinite medium calculated with some selected Skyrme functionals. In Sec.~\ref{sec:form:LR}, we briefly present the formalism of the LR theory and the modifications introduced by the extra terms present in the BSk family, we also discuss the stability properties of the infinite medium in the Landau limit and using the complete LR formalism.
In Sec.~\ref{neutrino:mfp}, we apply the LR formalism to calculate the NMFP at zero temperature. Finally, we present our conclusions in Sec.~\ref{conclusions}.


\begin{figure}[!h]
\begin{center}
        \includegraphics[clip,scale=0.34,angle=-90]{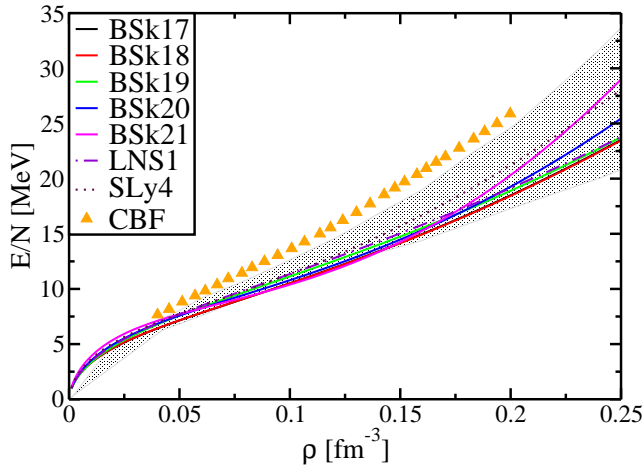}
\caption{(Color online) EoS in PNM for the the different functionals adopted in the text. The grey area represents the constraint obtained from Ref.~\cite{gan12}. The triangles correspond to the CBF calculations taken from Ref.~\cite{lov13}. }
\label{fig:eos}
\end{center}
\end{figure}

\section{Skyrme functional}\label{sec:form}

In its original article~\cite{sky56}, Skyrme proposed an effective pseudo-potential constructed as an expansion of the nuclear interaction in relative momenta, thus simulating finite-range effects in a zero-range interaction.
In its \emph{standard} form~\cite{cha97}, the Skyrme pseudo-potential is composed by a two-body central and a spin-orbit  term   plus a density dependent term to take into account three-body terms~\cite{vau72}.
The possibility of extending this \emph{standard} form has been recently investigated along two different ways: one following the spirit of the self-consistent mean field theory, where the major ingredient is an effective pseudo-potential and the beyond mean-field correlations are added afterwards~\cite{ben03}, and one following the spirit of the Density Functional Theory (DFT), where the building block is the functional which includes all correlations. 

The first approach has been followed by several groups who have studied the \emph{standard} pseudo-potential by adding to it the explicit tensor terms~\cite{les07}, higher order derivative terms~\cite{car08,dav13} or central 3-body terms~\cite{sad13}.
The DFT approach has been the guide-line for the development of the BSk family.
Higher order correlations are explicitly taken into account by the inclusion of 
a cranking approximation of the quadrupole correlations and
a phenomenological corrections for the Wigner energy.
Despite the remarkable good performances of the standard BSk functionals~\cite{gor09}, it has been found necessary to explore possible extensions of the functionals to satisfy astrophysical constraints especially the existence of two-solar mass neutron stars.

Starting from model BSk18~\cite{cha09}, two extra density-dependent terms have been added on top of a \emph{standard} Skyrme interaction. 
The addition of these extra terms provides flexibility to fulfill all available constraints coming from both finite nuclei and infinite matter.
In Appendix~\ref{app:functional}, we give the explicit expression of the functional used in the present article.

\begin{figure}
\begin{center}
        \includegraphics[clip,scale=0.34,angle=-90]{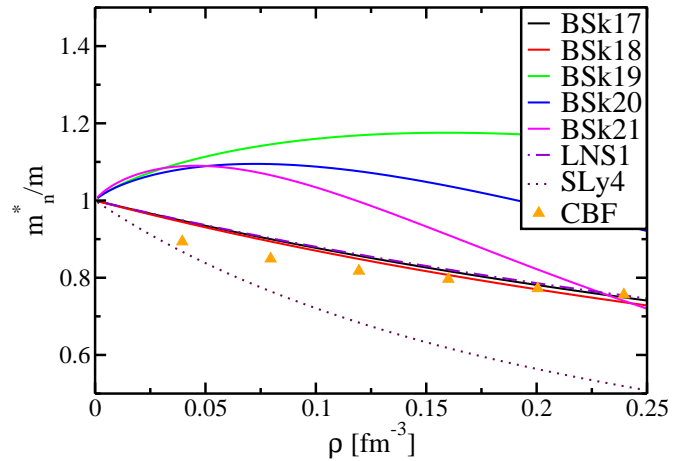}\\
      \caption{(Color online) Neutron effective mass in PNM as a function of the density, for the different functional. The triangles are obtained from Ref.~\cite{lov13} using the CBF method. }
\label{fig:meff}
\end{center}
\end{figure}

\noindent Starting from model BSk19~\cite{gor10}, the terms proportional to the square of the tensor current $J^{2}$ (we refer to~\cite{les07,cha10B} for a more detailed discussion) have been omitted, this means that all the coupling constants $C^{T}_{t}$ are set  equal to zero.
Doing this we loose the relation between coupling constants and coefficients of the pseudo-potential~\cite{rai11}, for such a reason we will use in the following only the functional formalism.

In the present article, we limit ourself to the the Brussels-Montreal functionals   BSk17~\cite{gor09}, BSk18~\cite{cha09}, BSk19-21~\cite{gor10}. 
All these models have been built to reproduce with high accuracy  all experimentally known masses of atomic nuclei~\cite{audi03} with a root-mean square deviation of $\sigma=0.58$~MeV.
The main differences arise from the properties of infinite nuclear matter,
and most particularly the EoS of PNM and the effective mass.
For the sake of comparison, we also consider two functionals that do not belong to the BSk family:~SLy4~\cite{cha97} and LNS1~\cite{gam11}. The first has been fitted considering both ground-state properties of some few selected even-even nuclei  and infinite matter properties; the latter has been constructed to reproduce Brueckner-Hartree-Fock (BHF) predictions in infinite matter~\cite{li08}.
In Fig.\ref{fig:eos}, we show the EoS for PNM for the different functionals. All the functionals give a very similar result up to $\rho\approx0.16$~fm$^{-3}$, where 
properties of nuclei still constrain the construction of the functionals.
Beyond this value, the different EoS show a different behavior, in particular BSk21  gives the stiffest EoS, while the softest is found for BSk18; the other functionals are more or less in between these two extreme cases.
On the same figure, we also report the results for the Correlated Basis Function (CBF) formalism~\cite{lov13} obtained with the Argonne $v_{6}^{\prime}$ two-body interaction plus the Urbana IX three-body term~\cite{pud95}, as discussed in Ref~\cite{lov14}. The shadow area represents the constraints obtained in Ref.~\cite{gan12} and it  gives an estimate of the uncertainties of these calculations, in particular concerning the choice of the three-body terms. It is important to recall that the functionals BSk17-19 have been fitted to reproduce the EoS in PNM of Friedman-Pandharipande~\cite{fri81}, while BSk20  to the EoS of Akmal and collaborators given in Ref.~\cite{akm98} and based on the calculations done with the Argonne interaction $v_{18}$+UIX+$\delta v$, where $\delta v $ is a relativistic boost.
Finally, BSk21  has been fitted to the results of Li and Schulze of Ref.~\cite{zi08}.
The SLy4 functional has been constrained to the EoS  of Wiringa \emph{et al.}~\cite{wir88}, and LNS1 has been fitted to the BHF results of Li \emph{et al.}~\cite{li08}.

Another important quantity used to characterize the ground state properties of PNM is the neutron effective mass. 
In Fig.\ref{fig:meff}, we show the density dependence of the neutron effective mass in PNM as obtained from different functionals. The additional density and momentum dependent terms in BSk19-21 modify the expression of the effective mass $m^{*}_{n}/m$ in PNM. In particular we notice that for the functional BSk21 the effective mass increases as a function of the density up to $m^{*}_{n}/m=1.1$ at $\rho_{n}\approx0.05$~fm$^{-3}$  and then it decreases to  $m^{*}_{n}/m=0.9$ at $\rho_{n}\approx0.16$~fm$^{-3}$. Such a behavior can be better appreciated for finite nuclei, as shown for example in Fig. 6 of Ref.~\cite{gor10}. This can not be reproduced by using a single density-dependent term.
In the same figure we also present the results obtained using the CBF method~\cite{lov14}. The effective mass obtained with this method is always lower than 1. This result is compatible with the neutron effective masses obtained with LNS1 and BSk17-18 functionals, while SLy4 gives relatively lower values.

\section{Linear Response Theory}
\label{sec:form:LR}

\subsection{Zero temperature}

In Sec.~\ref{sec:form}, we have characterized the infinite medium by defining some important quantities related to its ground state properties. In this section, we study the properties of its excited states.
The Random Phase Approximation (RPA) is a well suited tool for this analysis~\cite{Book:Fetter1971,Book:Ring1980}.
The method for calculating RPA response functions has been already presented in Refs. \cite{gar92,mar06,dav09} for symmetric nuclear matter (SNM)  in the case of a Skyrme pseudo-potential and then generalized to the case of an extended Skyrme functional in Ref.\cite{pas12} for SNM and PNM~\cite{pas12a}.
All the physical informations can be found in the dressed RPA propagator $G_{RPA}^{(\alpha)}(\mathbf{k}_1,q,\omega)$, where $(\alpha)$ stands for the quantum numbers of the system, $i.e$ in this case $\alpha=(S,M)$, where $S$ is the total spin and $M$ its projection, $\mathbf{k}_{1(2)}$ is the momentum of the incoming (outgoing) particle and $\mathbf{q},\omega$ are the transferred momentum and energy, respectively.  In the following, we will consider a system of natural units $\hbar=c=1$.
We refer to Refs.~\cite{pas13b,dav14b} for a more detailed discussion in the method.
The residual interaction between particles and holes (ph) for the functionals used in the present article, $V_{ph}^{(\alpha;\alpha')}$, takes the form %

\begin{eqnarray}\label{res:inter}
V_{ph}^{(\alpha;\alpha')}&=&\frac{1}{2}\delta_{\alpha,\alpha'} \left\{ \bar{W}_{1}^{(0)} +\bar{W}_{1}^{(1)} \text{\boldmath$\sigma$}_{a} \cdot\text{\boldmath$\sigma$}_{b}\right.\nonumber\\
& +&\left[ \bar{W}_{2}^{(0)} +\bar{W}_{2}^{(1)} \text{\boldmath$\sigma$}_{a} \cdot\text{\boldmath$\sigma$}_{b}\right]\left(\mathbf{k}_{1}-\mathbf{k}_{2}\right)^{2}\nonumber\\ 
&+&\left. \left[ \bar{W}_{3}^{(0)} +\bar{W}_{3}^{(1)} \text{\boldmath$\sigma$}_{a} \cdot\text{\boldmath$\sigma$}_{b}\right]\left[\mathbf{k}_{1}(\mathbf{q}+\mathbf{k}_{2})+\mathbf{k}_{2}(\mathbf{q}+\mathbf{k}_{1})\right]\right\}\nonumber\\ 
&-&i C^{\nabla J}\left(\text{\boldmath$\sigma$}_{a}+\text{\boldmath$\sigma$}_{b} \right)\cdot  \left[\mathbf{q} \times\mathbf{k}_{1}  -\mathbf{q}_{} \times \mathbf{k}_{2}  \right].
\end{eqnarray}

\noindent Compared to the residual interaction presented in Ref. \cite{pas12a}, we note the absence of the tensor terms ($C^{F}$) and an additional $\bar{W}^{(S)}_{3}$ term.
The latter has been already discussed in Ref.~\cite{gar92} in the case of a zero-range Skyrme-like interaction for liquid $^{3}$He~\cite{str84}.
It is interesting to stress that the residual interaction calculated for a Skyrme functional derived including  a three-body central zero-range pseudo-potential~\cite{sad13} gives rise to a similar structure, although the explicit value of the  $\bar{W}^{(S)}_{3}$ can be very different.

Despite the absence of a tensor term, it is important to note that there is still a coupling between the $S=0$ and $S=1$ channels and the removal of the $M$ degeneracy, due to the spin orbit term~\cite{pas12a}. Such a term enters in the final expression of the response function with a $q^{4}$ dependence, hence its effect on the response function is negligible only at very low values of the transferred momentum.
From the solution of the Bethe-Salpeter equations we obtain the dressed RPA propagator and the response function of the system as

\begin{equation}
\chi^{(\alpha)}(q,\omega)=g\int \frac{d^{3}\mathbf{k}}{(2\pi)^{3}}G_{RPA}^{(\alpha)} (\mathbf{k}_{},q,\omega) \quad .
\end{equation}

\noindent Here $g=2$ is the degeneracy of the system.
Since we  consider temperature effects, it is more relevant to  define the strength function as
\begin{equation}
S^{(\alpha)}(q,\omega) = -\frac{1}{\pi} {\rm Im} \chi^{(\alpha)}(q,\omega)\, ,
\end{equation}
which actually embeds all physical properties. The modification of these expressions for non-zero temperature has been discussed in Ref.\cite{dav14b}.
%
\begin{figure*}
\begin{center}
                 \includegraphics[clip,scale=0.36,angle=-90]{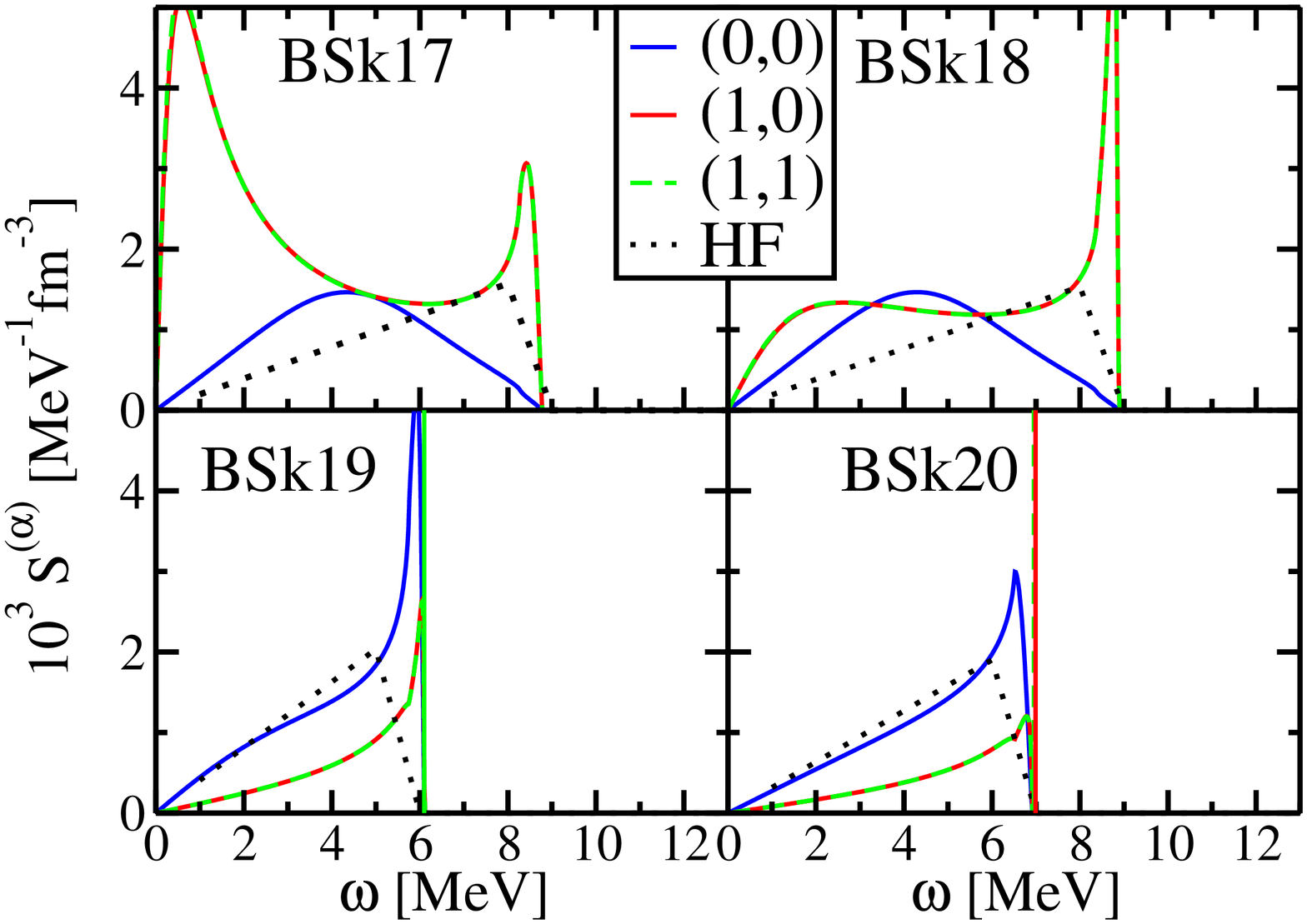}
                 \hspace{-2.94cm}
                 \includegraphics[clip,scale=0.36,angle=-90]{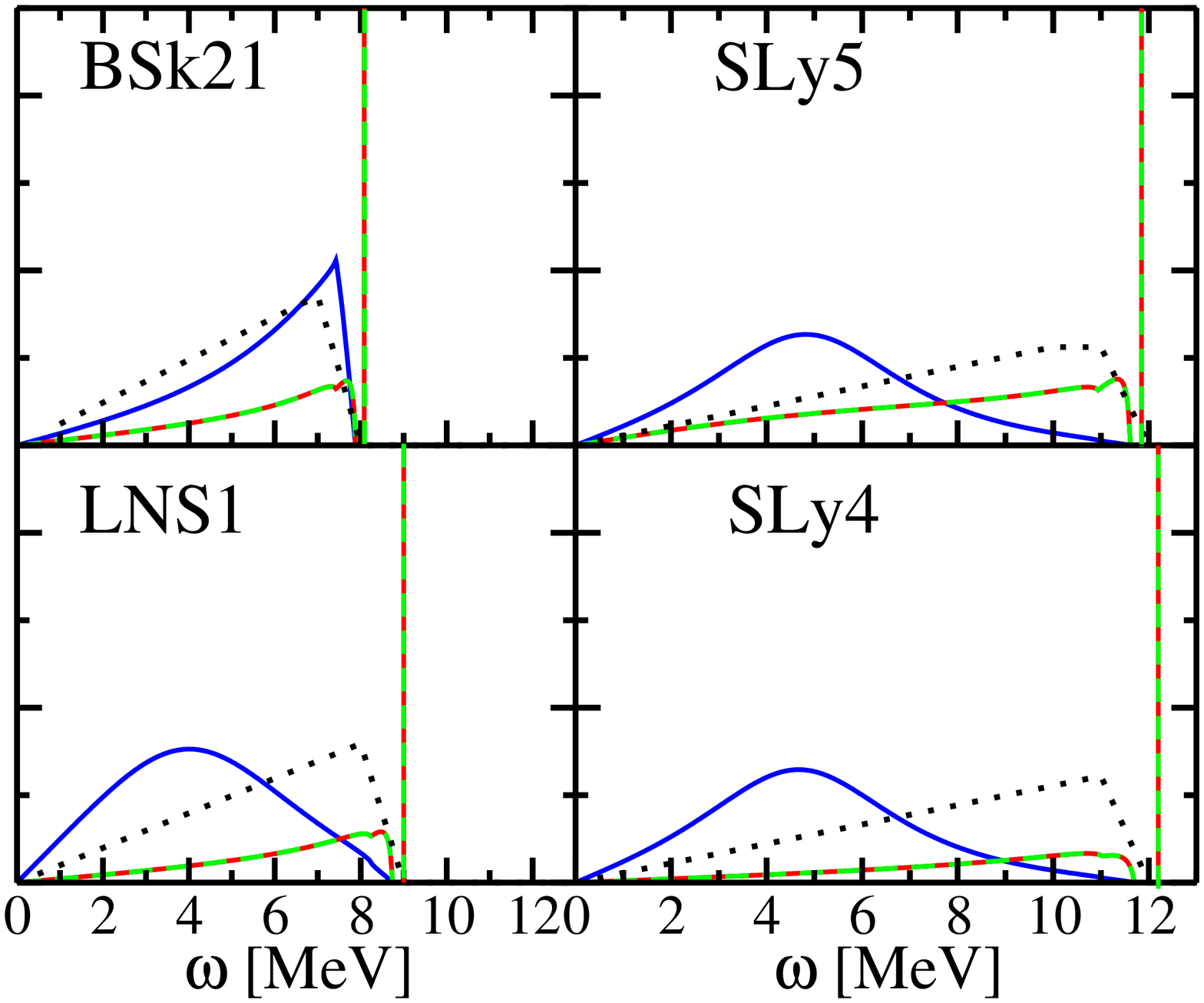}
\caption{(Color online) Strength function $S^{(S,M)}(q,\omega)$ calculated at $\rho=0.16 \;\text{fm}^{-3}$ and transferred momentum $q=0.1\; \text{fm}^{-1}$ for the different functionals used in the present study. The strength functions in the channels (1,0) and (1,1) are  degenerate. }
\label{fig:response}
\end{center}
\end{figure*}
%
In Fig.~\ref{fig:response}, we show the strength function at zero temperature and $\rho=0.16$~fm$^{-3}$ for the different functionals and for the transferred momentum $q=0.1$~fm$^{-1}$.
On the same figure we also present the Hartree-Fock (HF) strength function, $i.e.$ obtained by putting to zero the residual interaction.
Note that each response function is characterized by a maximum value of the energy transferred. This upper limit is directly related to the effective mass.
We observe that the two $S^{(1,M)}$ strength functions are almost degenerate in the spin projections $M=0,1$; indeed, at this value of transferred momentum the spin-orbit term gives negligible contributions due to its $q^{4}$ dependence.
Apart from BSk17, all the other functionals are repulsive in the spin channel $S=1$, and many of them manifest the appearance of a zero-sound mode~\cite{Book:Fetter1971}.
In this case the continuum part of the response function is suppressed in favor of this collective excitation.
To study the effect of the $J^{2}$ terms, we also consider the functional SLy5~\cite{cha97}. The latter predicts essentially the same properties for the ground-state of PNM as compared to SLy4, but it has non-zero contributions coming from the $J^{2}$ terms.
Comparing the two panels of Fig.\ref{fig:response}, we observe that the response function in the spin channel $S=0$ is not affected, while in the $S=1$ channel the response function is modified, SLy4 being more repulsive.
%
\begin{figure*}
\begin{center}
                \includegraphics[clip,scale=0.36,angle=-90]{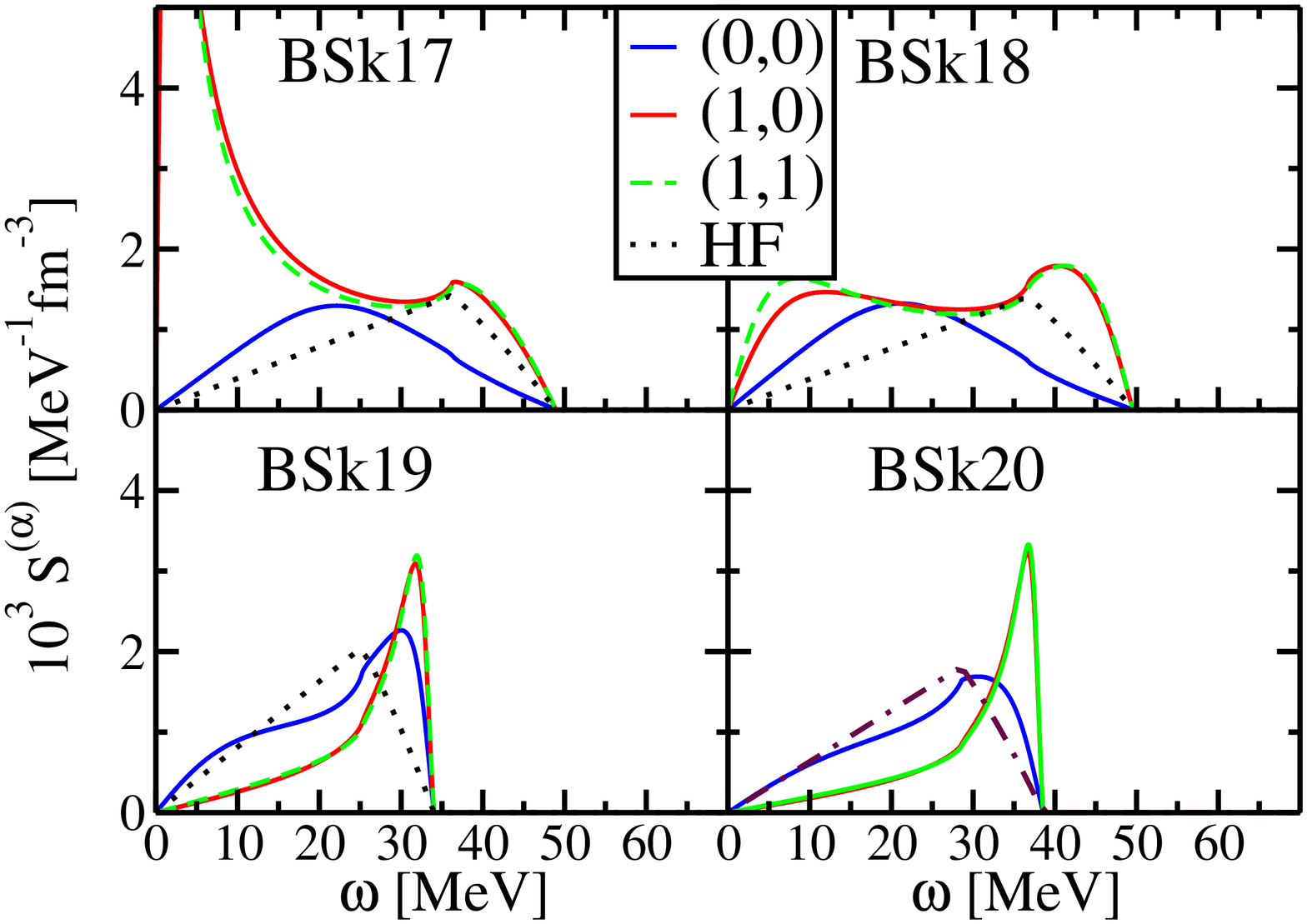}
                \hspace{-2.94cm}
                                \includegraphics[clip,scale=0.36,angle=-90]{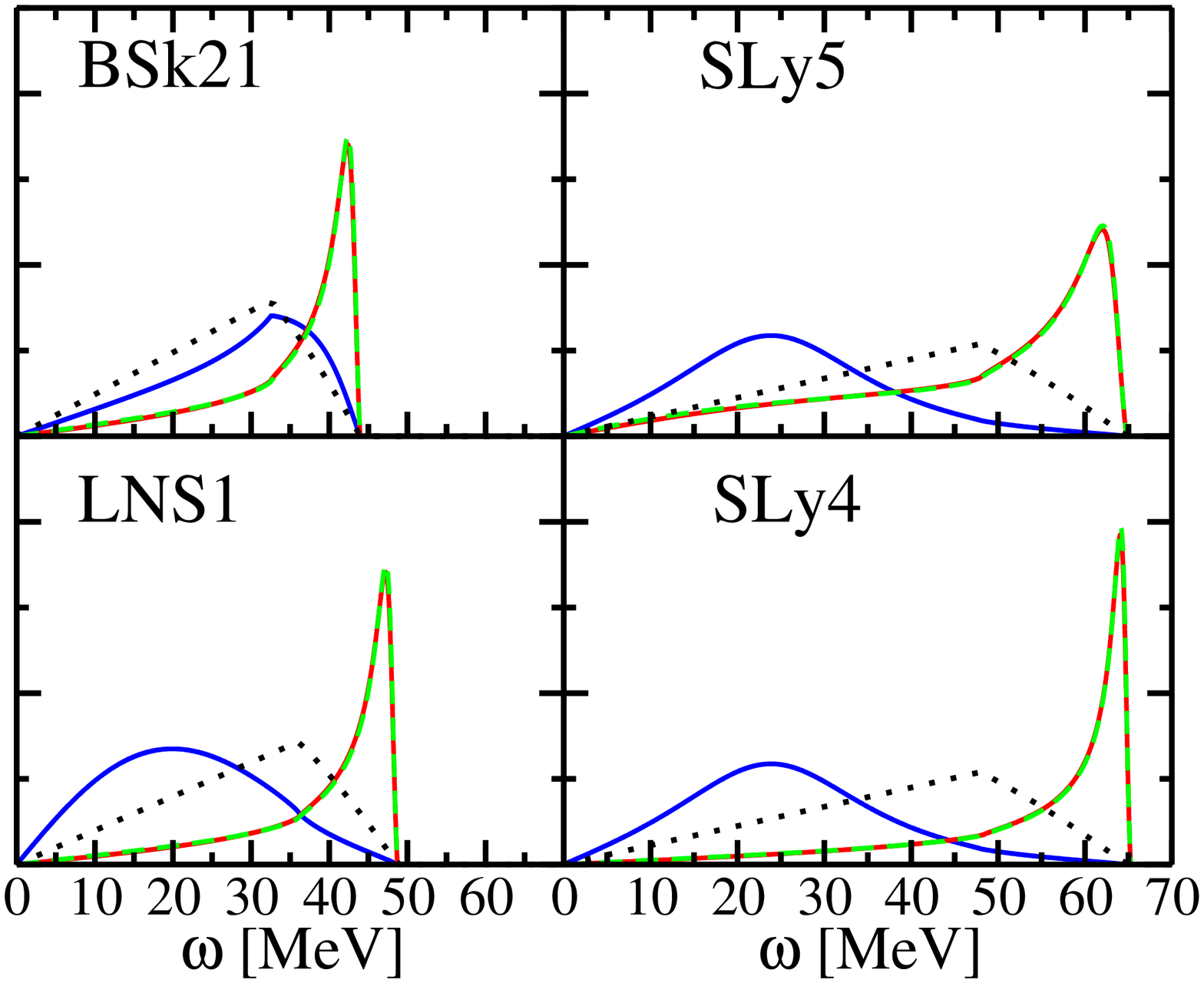}
\caption{(Color online) Same as Fig.\ref{fig:response}, but for $q=0.5$ fm$^{-1}$. }
\label{fig:response2}
\end{center}
\end{figure*}
%
In Fig.~\ref{fig:response2}, we show the response function at a higher transferred momentum $q=0.5$ fm$^{-1}$, but always at the same density. We observe that the zero sound mode is completely re-absorbed by the continuum part of the response function. The effect of the spin-orbit term starts to be visible for the BSk17-18 functionals where the two spin projections are no more degenerate, although the differences remain small at this value of the transferred momentum.
The BSk17 functional presents a strong accumulation of strength in the region of zero energy. This is the clear signal of the presence of an instability, as discussed in Sec.~\ref{sec:instability}.

\subsection{Thermal effects}\label{sec:T}

We now discuss how the temperature modifies the response function and eventually the NMFP.
The formalism of the LR theory at finite temperature has been already discussed in Refs.~\cite{dav14b,bra95,her96,her97}, and we refer to these articles for a more detailed discussion.

\begin{figure}
\begin{center}
                \includegraphics[clip,scale=0.34,angle=-90]{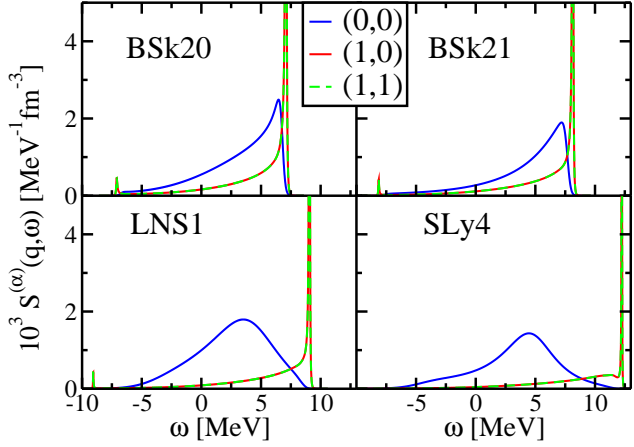}
\caption{(Color online) Response function for four selected functionals at $T=2$ MeV and $\rho=0.16$~fm$^{-3}$ and transferred momentum $q=0.1$ fm$^{-1}$. }
\label{fig:responseT}
\end{center}
\end{figure}

In Fig.~\ref{fig:responseT}, we show the response function at $T=2$~MeV for some selected functionals using the same values of the density and momentum as in Fig.\ref{fig:response}. 
It should be noticed that at zero temperature, the excitation energy is strictly positive, while at non-zero temperatures, negative energy excitations are  allowed.
Comparing with the results of Fig.~\ref{fig:response}, it can be seen that due to thermal fluctuations the zero sound mode is completely absorbed in the ph-continuum.
A small peak appears at the negative energy region, mirroring the zero sound mode. 


\subsection{Landau parameters}\label{inst:landau}

Since we consider excited states at a low exchanged momentum, it can be useful to briefly introduce the Landau theory of Fermi liquids~\cite{mig67} in PNM. 
Within this model, we consider only quasi-particles around the Fermi sphere which interact via a simple residual interaction of the form

\begin{eqnarray}
\label{landau-Vph}
 V_{ph}  &=&  \sum_{l} \bigg\{ f^{n}_{l} +  g^{n}_{l} \hat{\sigma}_1 \cdot \hat{\sigma}_2 \bigg\} P_{l} ( {\mathbf {\hat k}_1} \cdot {\mathbf {\hat k}_2}  )\nonumber\\
\end{eqnarray}
\noindent where  $f^{n}_{l},g_{l}^{n}  $ are the Landau parameters.
The Landau parameters can be related to the coupling constant of the Skyrme functional as
\begin{eqnarray}
f^{n}_{0}=N_{0}^{-1}F^{n}_{0}&=&\frac{1}{2}\bar{W}_{1L}^{(0)}+k_{F}^{2}\bar{W}_{2L}^{(0)}\\
f^{n}_{1}=N_{0}^{-1}F^{n}_{1}&=&-k_{F}^{2}\bar{W}_{2L}^{(0)}+k_{F}^{2}\bar{W}_{3L}^{(0)}
\end{eqnarray}
\noindent and similarly for the $g^{n}_{l}=N_{0}^{-1}G^{n}_{l}$. Here $N_{0}^{-1}=\frac{\hbar^{2}\pi^{2}}{2m^{*}k_{F}}$ is the inverse of the density of states at the Fermi 
surface~\cite{mig67}.
The coefficients $\bar{W}_{i=1,2,3;L}^{(S)}$  are obtained from the corresponding  $\bar{W}_{i=1,2,3}^{(S)}$ by substituting $q=0$ and are given in Appendix~\ref{w:pnm}.
In the case of standard Skyrme functionals that contain only second order derivatives, the only non-zero Landau parameters correspond to $l=0,1$ (S and P partial waves).
Including higher order derivatives, as for example the D-wave term, other non-zero terms would be possible, namely the $l=2$ term~\cite{dav13,dav14}.

\begin{figure}[ht]
\begin{center}
        \includegraphics[clip,scale=0.35,angle=-90]{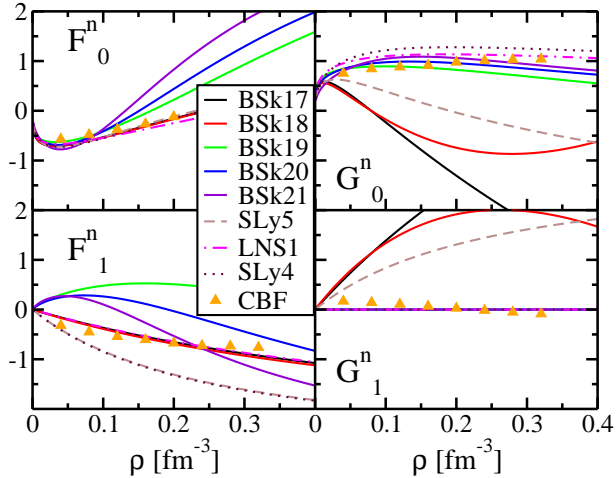}
\caption{(Color online) Landau parameters in PNM as a function of the neutron density. The triangles represent the results of CBF taken from Ref.~\cite{ben13}.}
\label{fig:landau:pnm}
\end{center}
\end{figure}

To guarantee the stability of the system the Landau parameters should fulfill some inequalities~\cite{mig67}, namely
\begin{eqnarray}\label{in:landau}
F_{l}>-(2l+1)
\end{eqnarray}
\noindent and similarly for $G_{l}$. Including tensor terms in the residual interaction requires additional conditions~\cite{bac79}. Otherwise, macroscopic quantities such as the compressibility and the static spin susceptibility becomes unphysical. 
Calculations based on standard Skyrme interactions have shown the presence of ferromagnetic instabilities~\cite{vid84,mar02,cha10B,isa04,cao10,nav13}. However, calculations of the static spin susceptibility based on \emph{ab-initio} methods~\cite{fan01,jac82,agu14}  satisfy these inequalities at least up to several times the saturation density.
Landau inequalities can be used as a constraint to impose the absence of instabilities in this density regime. This is the case of the latest BSk models starting from BSk18.
In Fig.~\ref{fig:landau:pnm}, we show the behavior of the Landau parameters in PNM as a function of the neutron density.
The functional BSk17 violates the inequality $G^{n}_{0}>-1$ at $\rho_{n}\approx0.16$~fm$^{-3}$ leading to a ground state of polarized neutron matter, as already pointed out in Ref.~\cite{cha09}. 
The SLy5 functional also violates such an inequality at $\rho_{n}\approx0.6$~fm$^{-3}$, but this value is sufficiently high not to affect the calculations of NMFP presented in the following sections.
%
  %
%

Landau parameters are also related to specific properties of infinite nuclear matter, such as the EoS, the effective mass and the incompressibility~\cite{pas12}.
Starting from BSk19, the BSk functionals have no $J^{2}$ terms, thus the corresponding Landau parameter $G^{n}_{1}$ is identically zero, similarly to LNS1 and SLy4.
As seen in Fig.~\ref{fig:landau:pnm}, the Landau parameters obtained with LNS1 gives rather similar results as those obtained with CBF~\cite{ben13}. This is not surprising, since LNS1 has been explicitly fitted to reproduce some infinite matter properties obtained with {\it ab-initio} methods.
A complete discussion over Landau parameters goes beyond the scope of this study, but it is important to remark that there is no general consensus over the magnitude or even the sign of some Landau parameters (see for example the discussion in Ref.~\cite{pas13b}).

\begin{figure*}[!h]
\begin{center}
      \includegraphics[clip,scale=0.3,angle=-90]{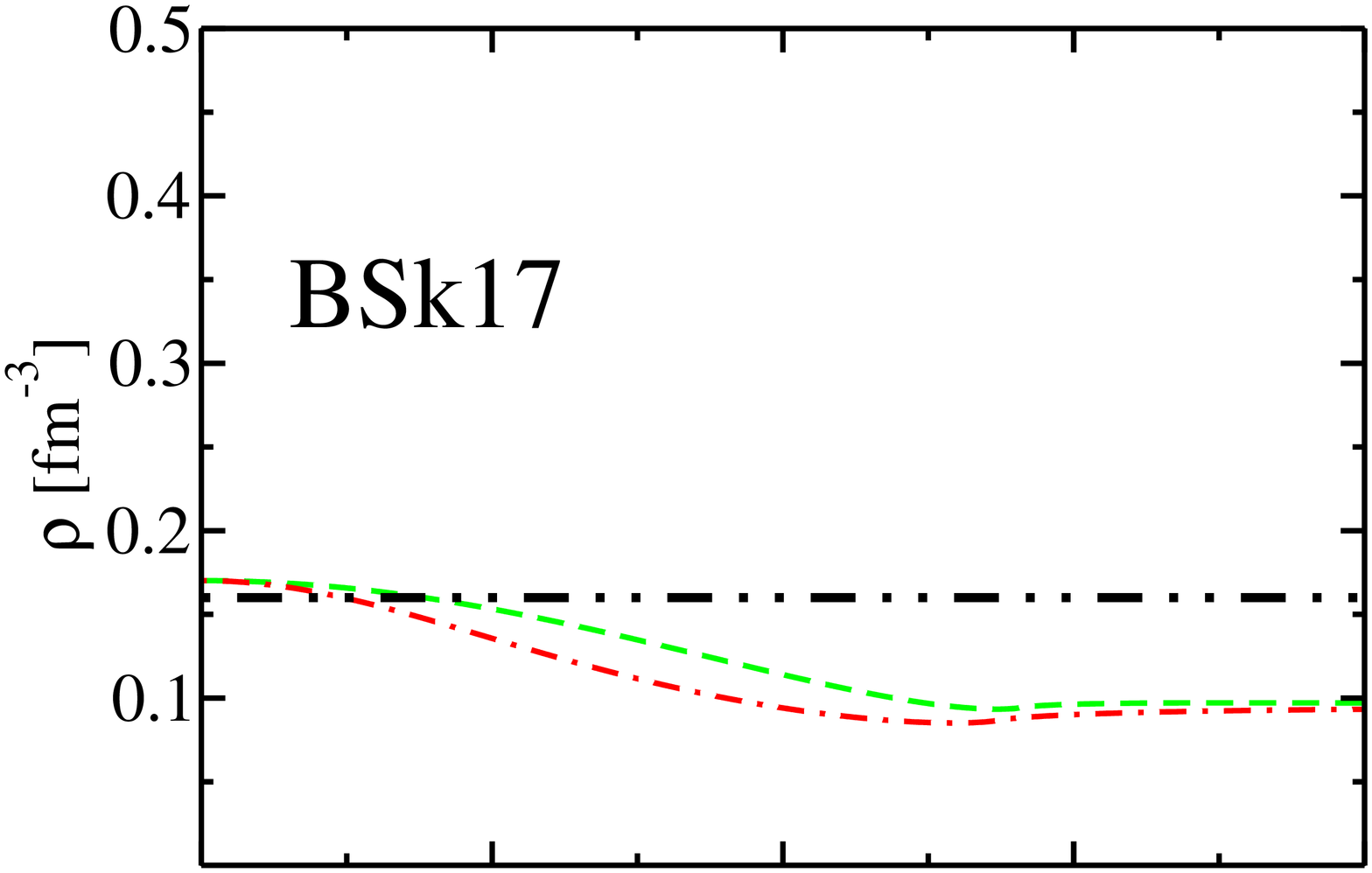}
                      \hspace{-2.3cm}
      \includegraphics[clip,scale=0.3,angle=-90]{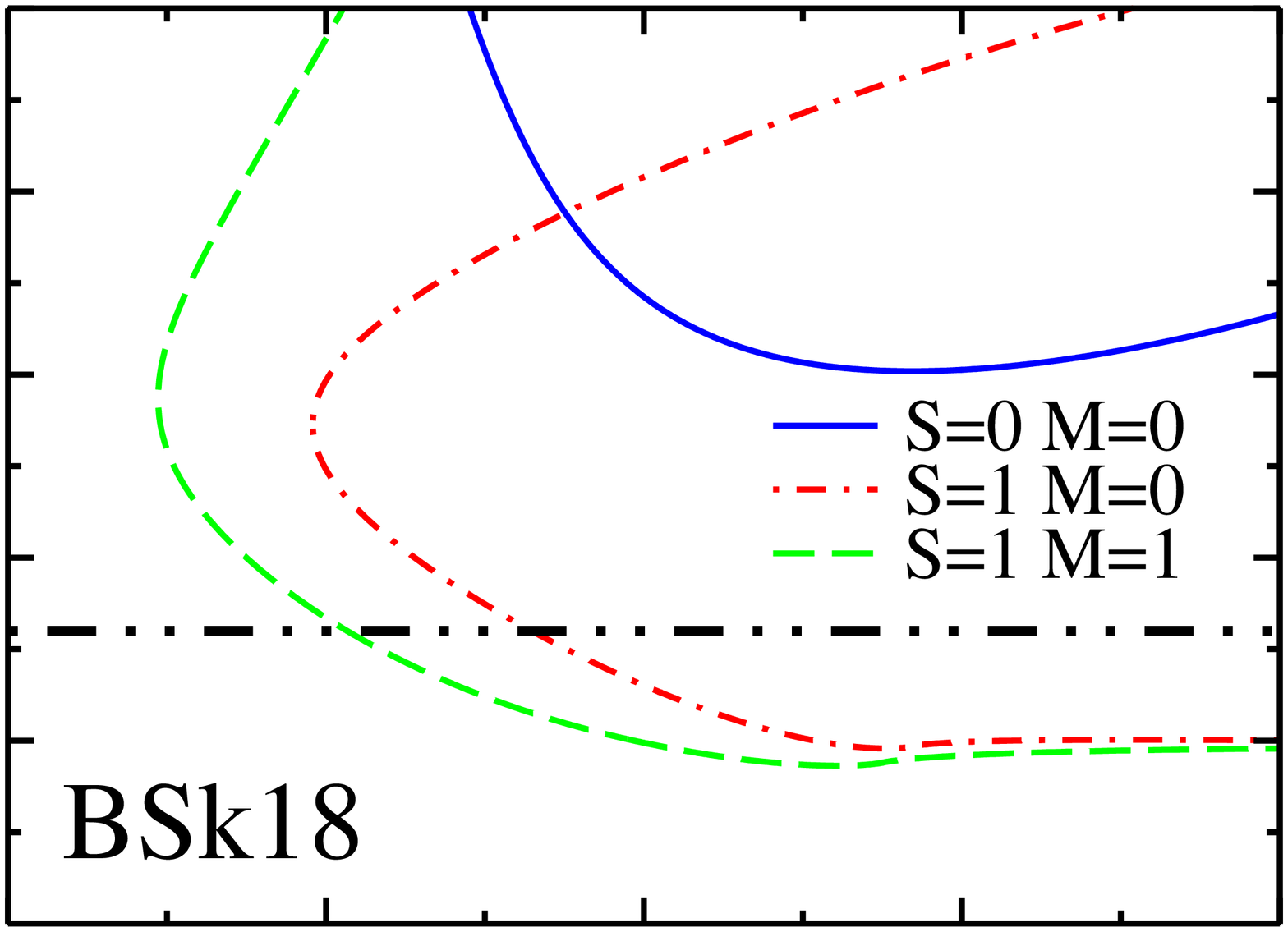}\\
            \vspace{-2cm}
      \includegraphics[clip,scale=0.3,angle=-90]{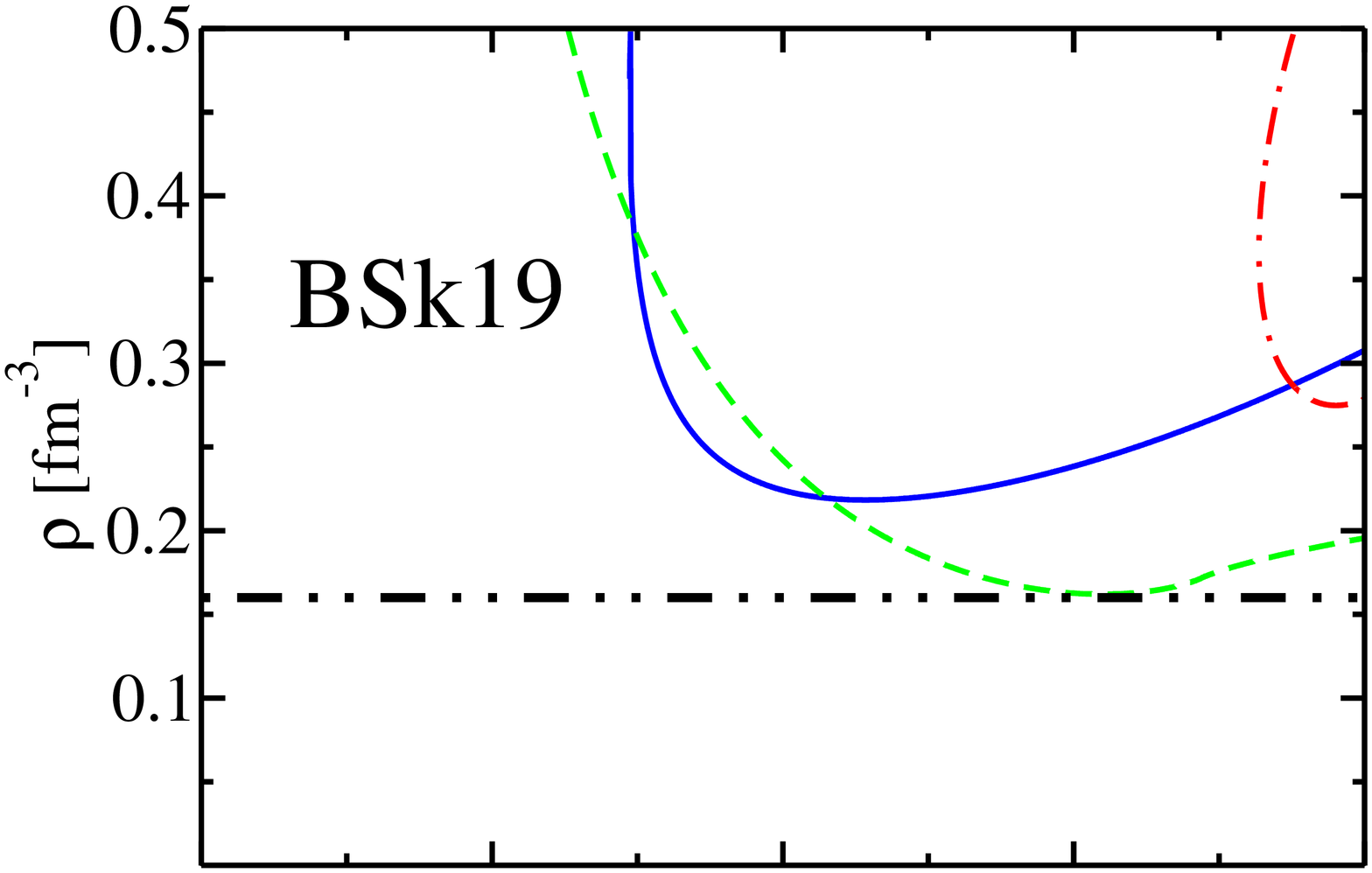}
                      \hspace{-2.3cm}
      \includegraphics[clip,scale=0.3,angle=-90]{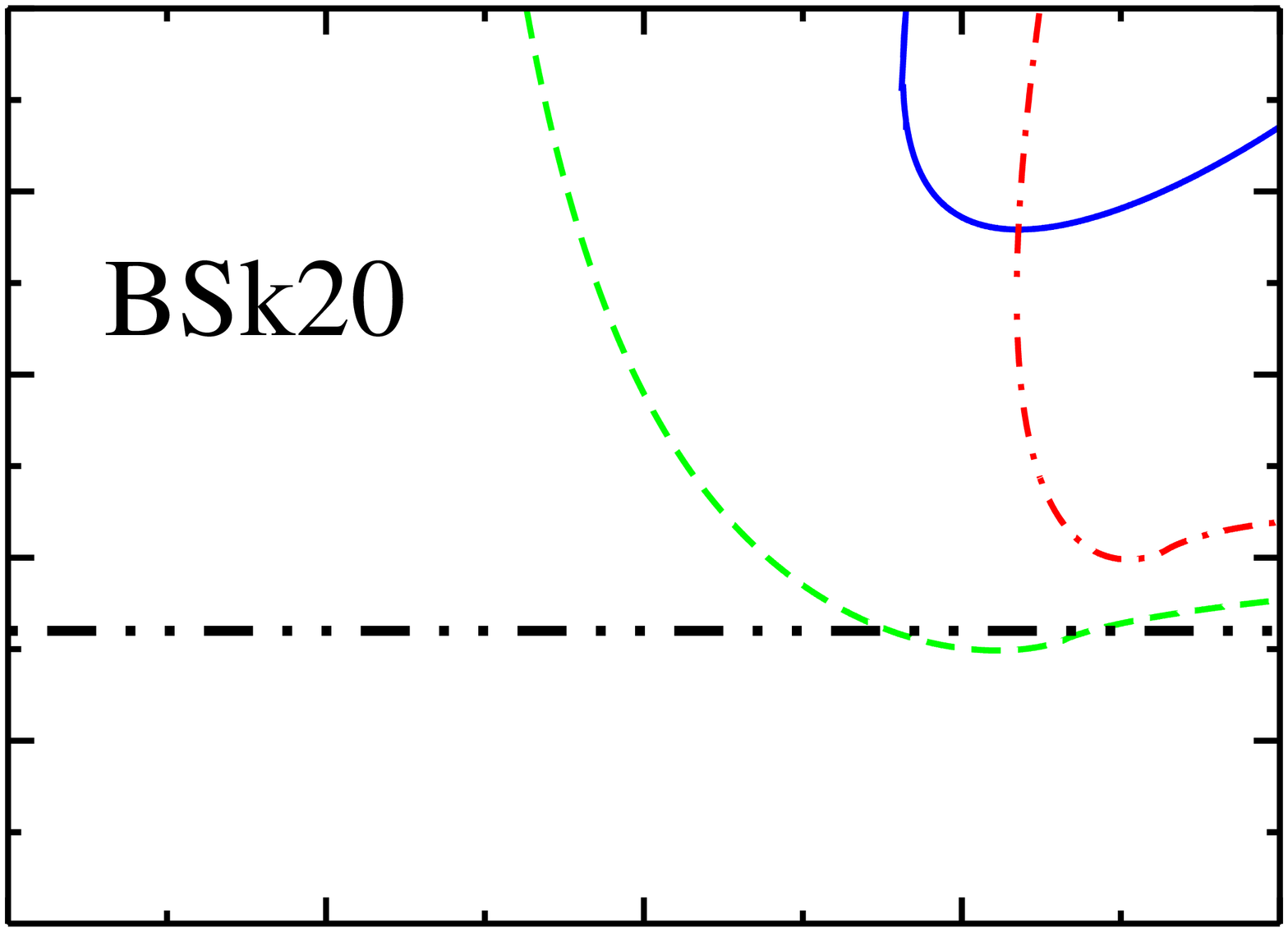}\\
      \vspace{-2cm}
     \includegraphics[clip,scale=0.3,angle=-90]{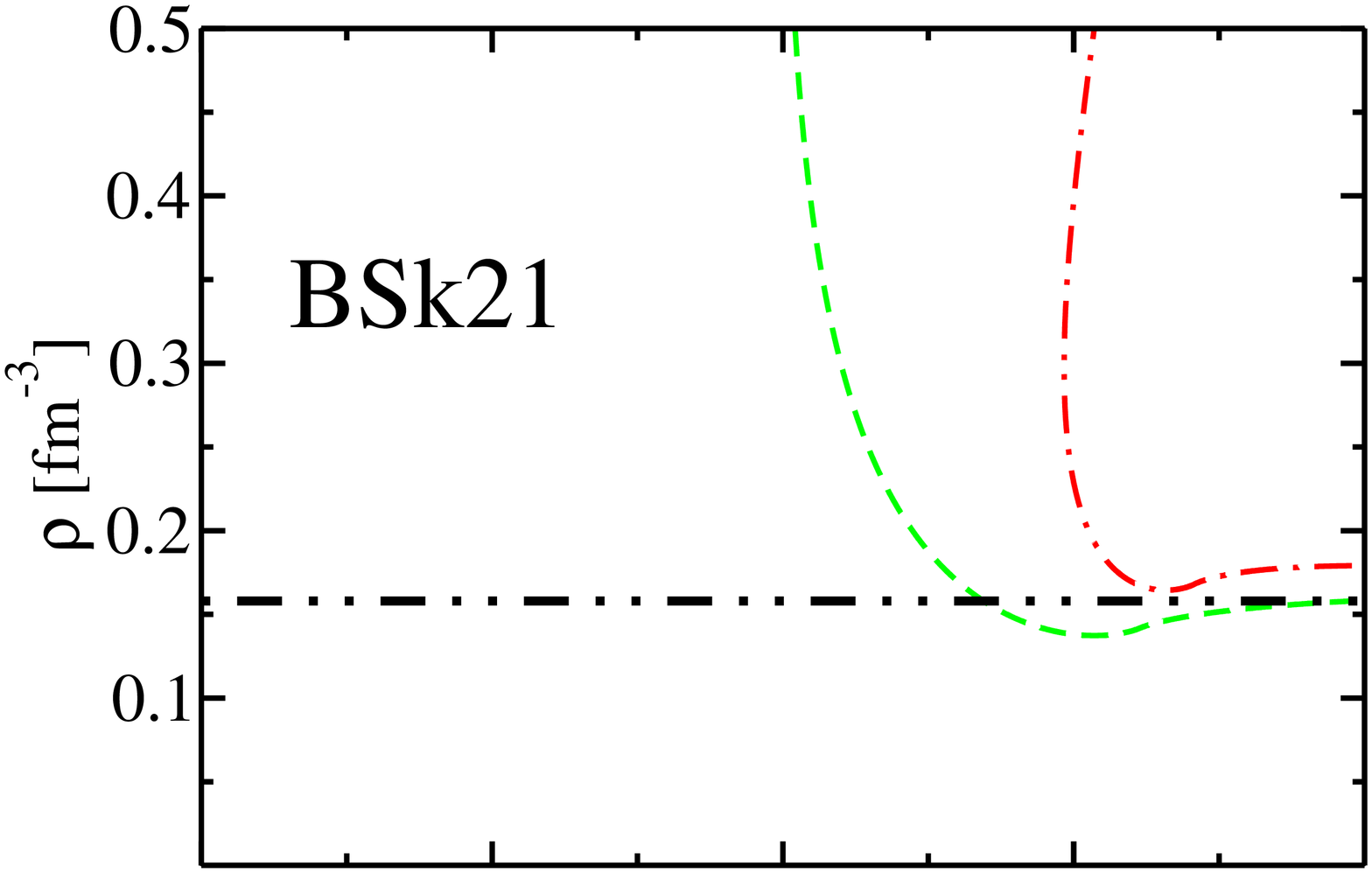}
            \hspace{-2.3cm}
         \includegraphics[clip,scale=0.3,angle=-90]{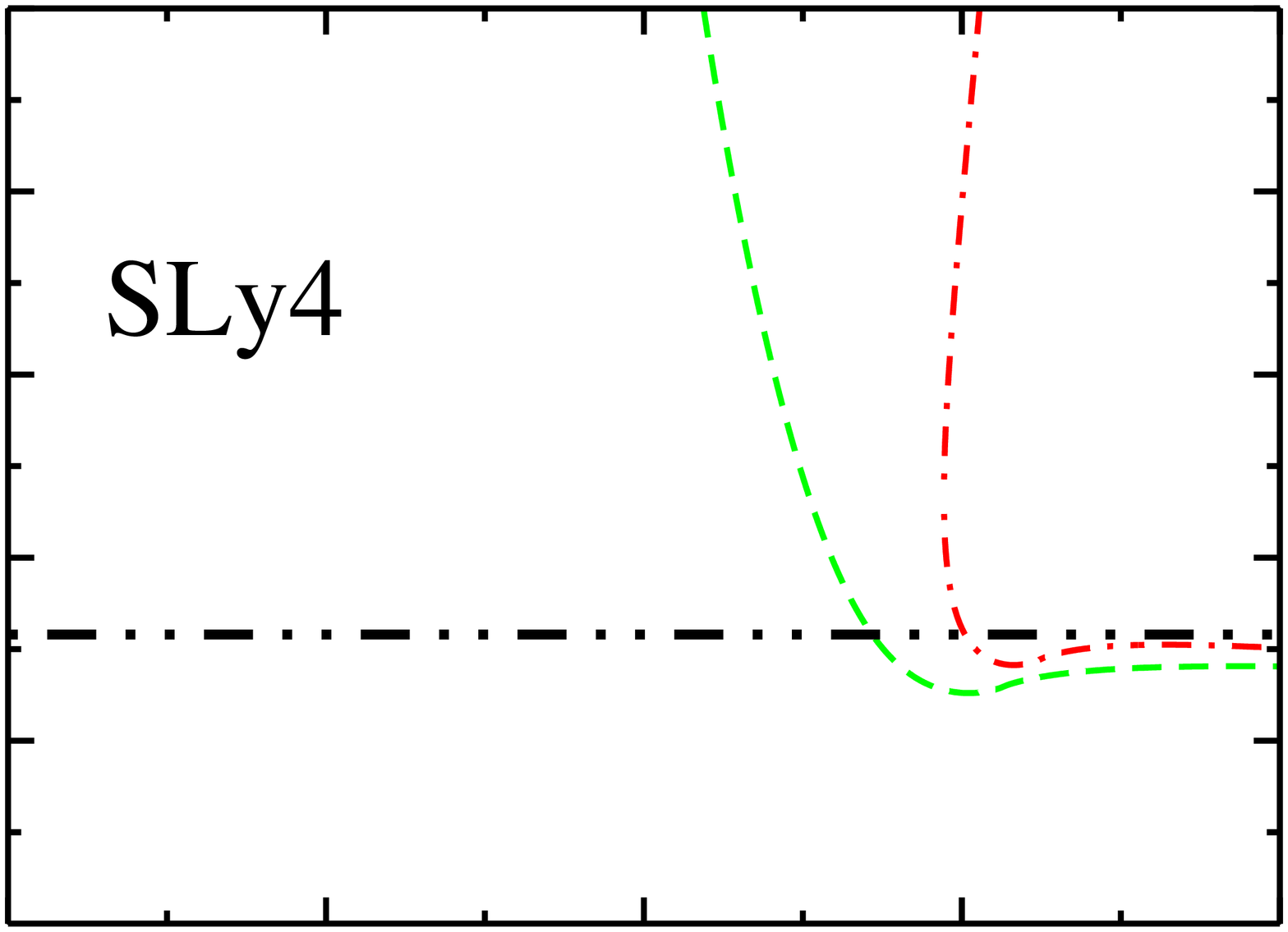}\\
                              \vspace{-2cm}
             \includegraphics[clip,scale=0.3,angle=-90]{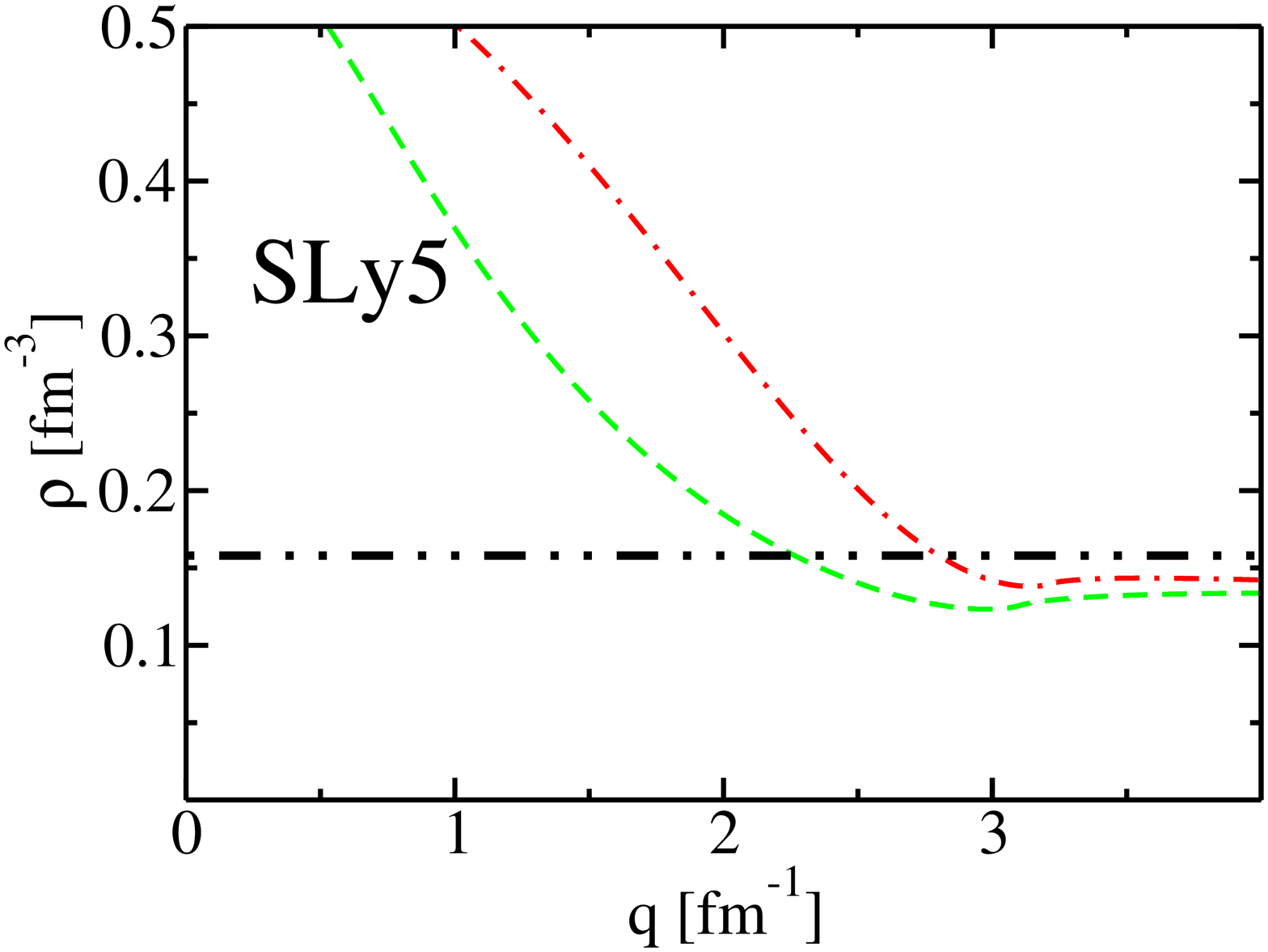}
                      \hspace{-2.3cm}
      \includegraphics[clip,scale=0.3,angle=-90]{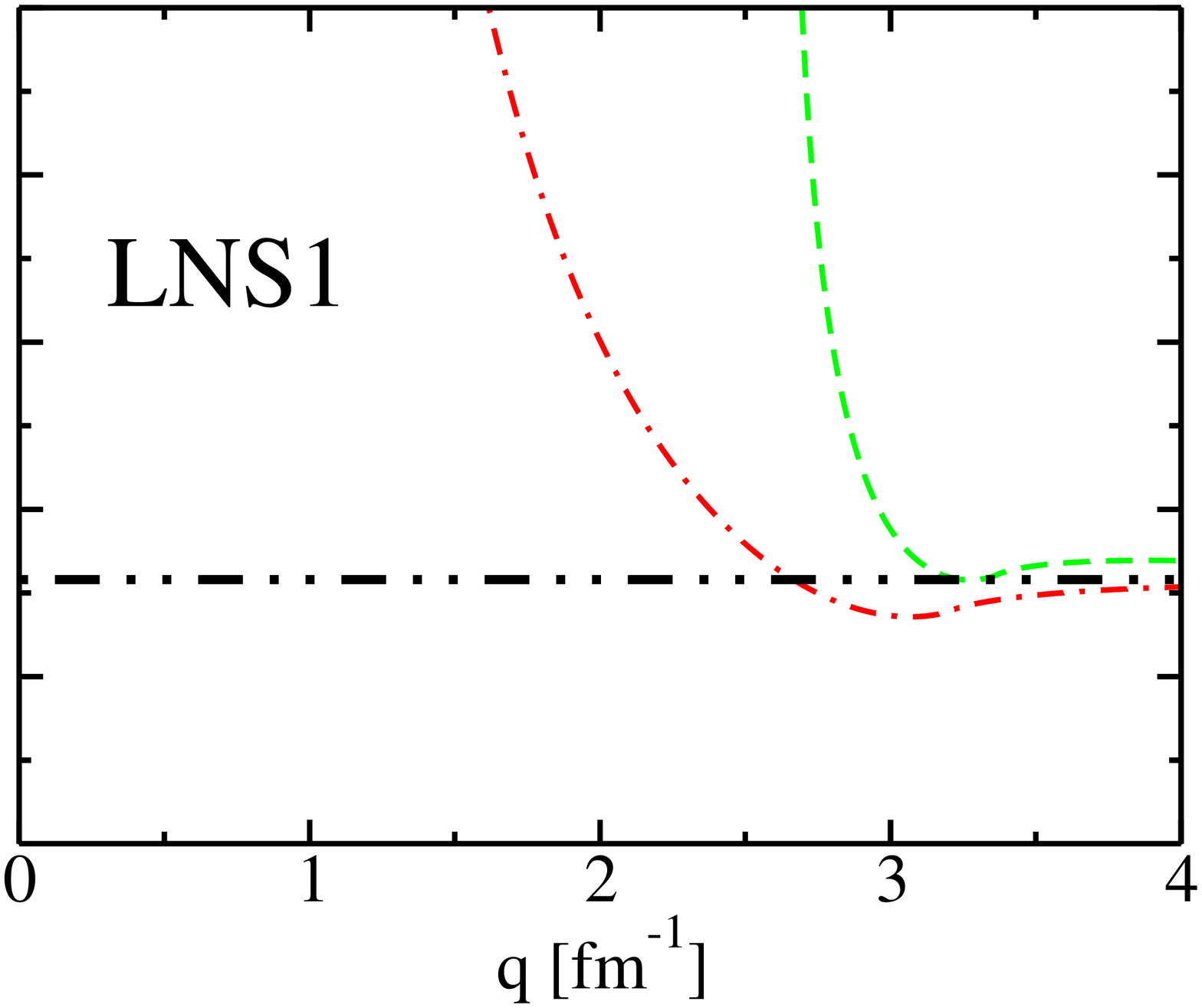}
\caption{(Color online) Critical densities in PNM for different Skyrme functionals as a function of the transferred momentum $q$. The horizontal dashed-dotted line represents $\rho=0.16$ fm$^{-3}$.}
\label{fig:critic:pnm}
\end{center}
\end{figure*}


%
%
%
\subsection{Instabilities in PNM}\label{sec:instability}
The instabilities discussed in the previous Section refer to the so-called long wave-length limit. 
Other instabilities could also appear, in connection to the properties of the $ph$ interaction away from the Fermi surface.
Such an instability consists in a phase transition of finite-size domain of length $\lambda\approx2\pi/q$ where $q$ is the momentum exchanged with an external probe.
Within the RPA formalism, such an instability manifests through the appearance of a mode at zero energy and infinite strength.
An example of such an instability can be seen in Fig.~\ref{fig:response2} for the BSk17 functional.

To detect these poles we  solve the  equation
\begin{equation}
1/\chi^{(\alpha)}(\omega=0,q)=0\,,
\end{equation}

\noindent for different values of the density  of the system $\rho$ and the transfer momentum $q$. 
In Fig.~\ref{fig:critic:pnm}, we show the position of the instabilities for the different functionals adopted in the article.
The functional BSk17 presents an instability at $q=0$ in the spin channel which is related to the violation of the Landau inequality as shown in Fig.\ref{fig:landau:pnm}. All the functionals present finite-size instabilities in the spin channel, this means that PNM could develop domains polarized in spin.

As previously discussed, realistic calculations predicts PNM free from instabilities in the low momentum regime up to very high density regions. These results have been used as constraints during the construction of some Skyrme functionals~\cite{mar02,cha09}.
Only recently, finite size instabilities, though in symmetric nuclear matter, have been investigated and in particular their relations with instabilities in finite nuclei~\cite{les06,hel13}.
The nature of these instabilities can vary according to the channel in which they manifest~\cite{fra12,sch10,hel12}. 
Concerning PNM, a possible constraint could come from some \emph{ab-initio} calculation of neutron droplets~\cite{gan11}. In Ref.~\cite{kor13}, calculations of neutron droplets with Skyrme functionals have been presented. It has  been shown that  instabilities detected in PNM can also manifest in such  calculations.
A more systematic investigation is required, but this method could be useful at least to determine if the presence of finite-size instabilities in PNM in low density regions is a pathology or not of the Skyrme functional.

\section{Neutrino mean free path}\label{neutrino:mfp}

Among the different processes involving neutrinos during the evolution of a neutron star, the scattering of thermal neutrinos over layers of nuclear matter plays an important role. We refer to Ref.~\cite{red98b} for a more detailed discussion.
In the present section we adopt the formalism of the LR theory at finite temperature to study the sensitivity to the nuclear interaction.

The neutrino mean free path  due to scattering inside neutron matter in the case of non-degenerate neutrinos can be written as $\lambda=(\sigma\rho)^{-1}$, where $\sigma$ is the total cross section for the neutral current  reaction $\nu+n\longrightarrow \nu'+n'$.
The general formula of the double differential cross section reads 
\begin{widetext}
\begin{eqnarray}\label{diff:cross}
\frac{d^{2}\sigma (E_{\nu})}{d\Omega_{k'}d\omega}&=&\frac{G_{F}^{2}E_{\nu}^{2}}{16\rho \pi^{2}}\left\{ (1+\cos\theta ) S^{(0,0)}(q,\omega,T) +g_{A}^{2}\left[ \frac{2(E_{\nu'}\cos\theta-E_{\nu})(E_{\nu'}-E_{\nu}\cos\theta)}{q^{2}}+1-\cos\theta\right]S^{(1,0)}(q,\omega,T)\right.\nonumber\\
&+&{2g^{2}_{A}}\left. \left[ \frac{E_{\nu}E_{\nu'}}{q^{2}}\sin^{2}\theta+1-\cos\theta\right]S^{(1,1)}(q,\omega,T)\right\}\,,
\end{eqnarray}
\end{widetext}

\begin{figure}
\begin{center}
      \includegraphics[clip,scale=0.35,angle=-90]{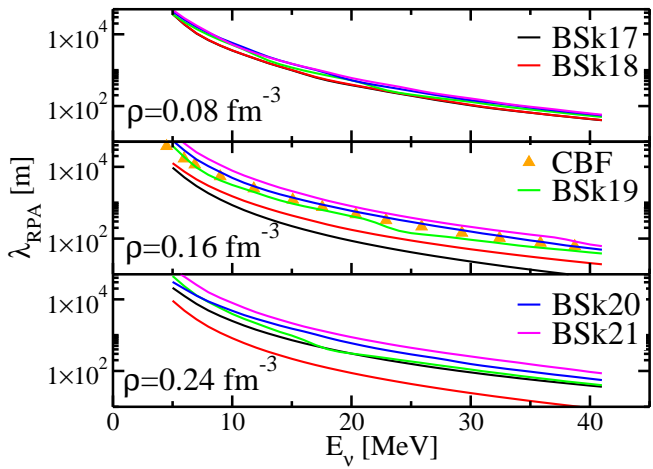}
      
\caption{(Color online) NMFP $\lambda_{RPA}$ at $\rho=0.16$ fm$^{-3}$ and zero temperature for different BSk  functionals. The triangles represent the result obtained in Refs.~\cite{ben13,lov14} using CBF method.}
\label{fig:nmfp:bsk}
\end{center}
\end{figure}

\noindent where $G_{F}$ is the weak coupling constant and $g_{A}$ is the axial charge of the nucleon. $E_{\nu(\nu')}$ is the energy of the incoming (outcoming) neutrino, while $\theta$ is the scattering angle between them.
The quantities $\omega=E_{\nu}-E_{\nu'}$ and $\mathbf{q}=\mathbf{k}-\mathbf{k}'$ are the energy and momentum transfer in the reaction.
As already discussed in Refs.~\cite{mar09,mar10,pas12a}, the cross section is dominated by the spin transverse response $(S=1,M=1)$.

In Fig.~\ref{fig:nmfp:bsk}, we show the NMFP $\lambda_{RPA}$ at  three different densities and at zero temperature for the different BSk functionals.
On the same figure, we report the results of Refs.~\cite{ben13,lov14} using the CBF method. The NMFP decreases quickly for increasing values of incoming neutrino energy $E_{\nu}$.
The functional BSk17 gives a  small mean free path at $E_{\nu}\approx$40 MeV and $\rho=0.16$ fm$^{-3}$, compared to the other functionals the reduction is much more pronounced. This is due to the presence of finite-size instabilities in this density region (cfr. Fig.\ref{fig:critic:pnm}). A similar behavior is shown by BSk18, but at density $\rho=0.24$ fm$^{-3}$.
The available results with CBF at $\rho=0.16$ fm$^{-3}$ are compatible with the results of the BSk19-21 functionals. This can be understood by looking at the properties of the Landau parameters shown in Fig.\ref{fig:landau:pnm}. In this case the Landau parameters $G_{0,1}^{n}$ obtained with CBF are very close to those of the BSk19-21 functionals for this given value of the density.
Although we recall that for CBF we have an infinite series of Landau parameters, in Ref.~\cite{pas13b} it has been shown that the response function $S^{(\alpha)}(q,\omega)$ converges quickly as a function of $l$ and the inclusion of the first partial waves up to $l=2$ is sufficient to obtain a well converged response function.
%
\begin{figure}
\begin{center}
            \includegraphics[clip,scale=0.35,angle=-90]{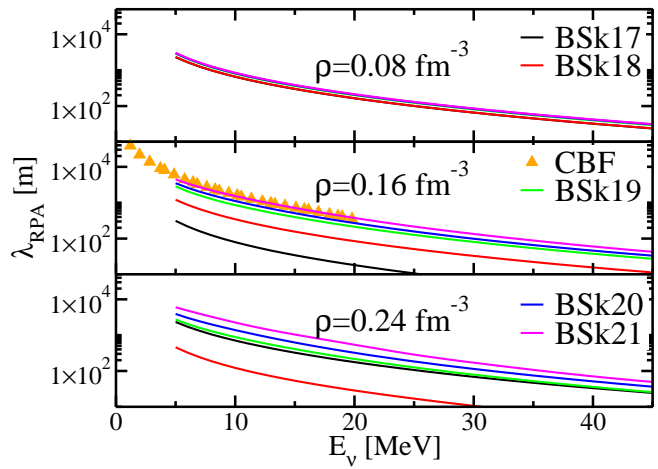}
\caption{(Color online) Same as Fig.\ref{fig:nmfp:bsk}, but for T=2 MeV.}
\label{fig:nmfp:bsk:T2}
\end{center}
\end{figure}
%
In Fig.\ref{fig:nmfp:bsk:T2}, we show the equivalent results at a temperature T=2 MeV.
The  temperature affects directly the response function as already shown in Fig.\ref{fig:responseT}, by allowing also for negative energy excitations and changing the widths of the peaks. It also modifies the integration limits imposed in the calculation of $\lambda$ as discussed in Ref.~\cite{iwa82}.
Again we observe that the CBF are in reasonable agreement with the results obtained using the BSk19-21 functionals.

\begin{figure}
\begin{center}
      \includegraphics[clip,scale=0.36,angle=-90]{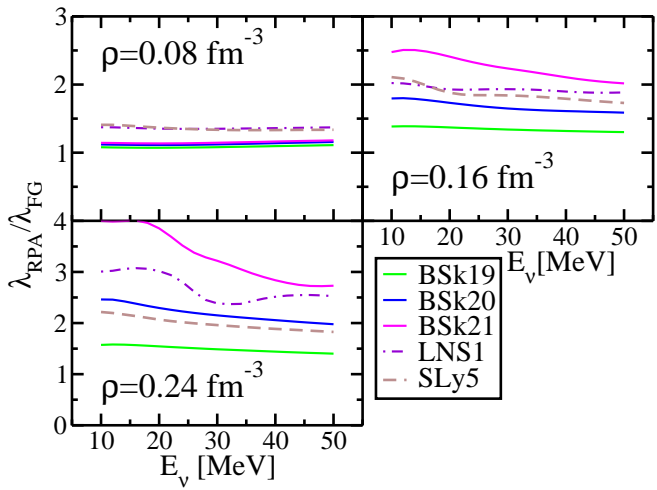}
      
\caption{(Color online) Ratio of the NMFP obtained with the full calculation $\lambda_{RPA}$ to the one based on the simple Fermi Gas $\lambda_{FG}$ as a function of the neutrino energy at $T=2$ MeV.}
\label{fig:nmfp:ratioT2}
\end{center}
\end{figure}

To better separate the effects of neutrino kinematics from the effects of the interaction we present, in Fig.\ref{fig:nmfp:ratioT2}, the ratio of the NMFP calculated using the LR theory and the complete  interaction $\lambda_{RPA}$ to the results for the Fermi Gas (FG) $\lambda_{FG}$.
As previously discussed, we have removed the BSk17-18 functionals since they present some instabilities in the low momentum region.

In the low-density regime, the BSk19-21 functionals clearly predict similar results, $i.e.$ the presence of the residual  interaction leads to a small increase by typically $\approx15$\% of the NMFP compared to the FG case. The functionals SLy5 and LNS1 on the contrary predicts a higher ratio close to 1.4.
At higher density, major differences are found with ratios ranging between 1.5 for BSk19 and 4 for BSk21. These variations can be related to the different EoS which differ significantly at high densities, as illustrated in Fig.\ref{fig:eos}. Finally, we note that the BSk20 results are quite similar to those obtained with SLy5 in all the three density regimes.
Despite their differences in the effective mass and Landau parameters prediction, the NMFP turns out to be rather similar.
\begin{figure}
\begin{center}
      \includegraphics[clip,scale=0.35,angle=-90]{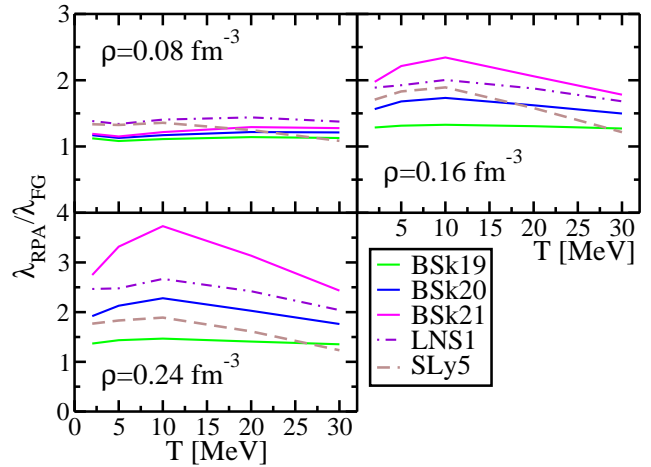}
      
\caption{(Color online) Ratio of the RPA to the FG  NMFP as a function of the temperature of the system at different densities, as indicated in the panels,  and for a fixed value of incoming neutrino energy $E_{\nu}=60$ MeV .}
\label{fig:nmfp:bsk3}
\end{center}
\end{figure}
In Fig.~\ref{fig:nmfp:bsk3}, we show the evolution of the NMFP as a function of the temperature for a fixed value of the incoming neutrino energy, $E_{\nu}=60$ MeV and for different values of the densities as indicated in the different panels.
A  weak dependence on the temperature is found for the BSk19-20 functionals, in contrast to BSk21 which shows a quite remarkable temperature dependence in the high density region.
At $T=30$ MeV the NMFP predicted by BSk21 at $\rho=0.24$ fm$^{-3}$ is still a factor of 2 larger than the one for the non-interacting FG. In this respect, a non-negligible effect in the thermal evolution of massive stars could be expected at the end of their thermonuclear burning.

\begin{figure}
\begin{center}
      \includegraphics[clip,scale=0.35,angle=-90]{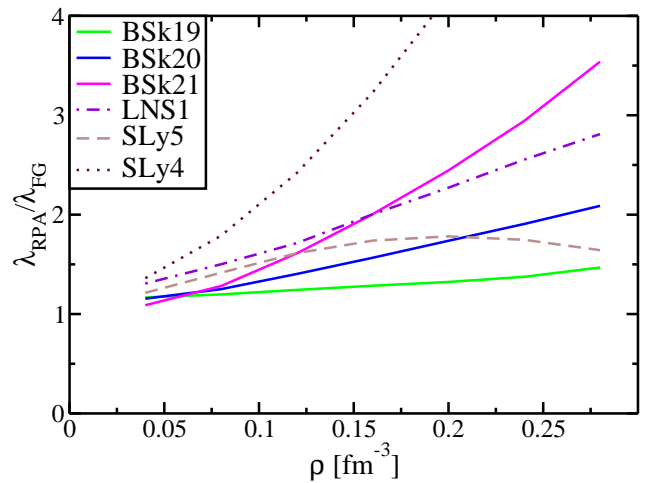}
      
\caption{(Color online)Ratio of the of the T=0 NMFP $\lambda_{RPA}/\lambda_{FG}$  for a fixed neutrino energy $E_{\nu}=60$ MeV.}
\label{fig:nmfp:bsk4}
\end{center}
\end{figure}

In Fig.~\ref{fig:nmfp:bsk4}, we show the evolution of the T=0 NMFP as a function of the density of the system for a fixed neutrino energy of $E_{\nu}=60$ MeV.
We clearly observe that for all the functionals the ratio of NMFP increases with the density, apart from SLy5 which shows a different behavior.
The presence of an explicit tensor terms can strongly affect the mean free path;  its evolution  as a function of the density can be more complex than the one shown here, see for example Fig.11 of Ref.~\cite{pas12a}.
In the present calculation, the main difference between the functionals SLy5 and SLy4 is the contribution (or not) of the tensor term to the central part of the residual interaction. These functionals have been obtained using the same fitting protocol~\cite{cha97}, and they have essentially the same effective mass and EoS.
The absence of the so-called $J^{2}$ terms in SLy4 makes the NMFP larger compared to the one of SLy5, moreover the evolution as a function of the density of the system is opposite: SLy4  rapidly increases while SLy5 gives a flatter behavior.  

%
%

\section{Conclusions}\label{conclusions}

In this article we have derived the formalism of the LR theory for the BSk family of Skyrme functionals. These functionals present a non-standard density dependence, so that we have generalized the results already presented in Ref.~\cite{pas12a} to take into account such extra terms.
The interest of these functionals is that they have been built using constraints on both finite nuclei and infinite matter properties and are thus well suited for calculations of astrophysical interest.
The LR theory is an important tool since it allows to test through simple and rapid calculations the residual interaction in the different spin channels and eventually to detect finite size instabilities.
The latter have been recently investigated in a systematic way~\cite{hel13} providing a simple criterion  based on the position of poles in the infinite medium to avoid them. 

In the present article, we have also presented a direct astrophysical application of the calculations of the nuclear response function, namely the calculations  of the NMFP at different densities and temperature of the system. Although such calculations can be considered still schematic since they do not take into account properly the stellar medium ($i.e.$ the eventual proton/electron fraction), we can clearly observe a variety of trends and behaviors of the different functionals.
The vector part of the residual interaction is quite difficult to constrain by looking at properties of finite nuclei, while in these calculations it plays a major role. 
A systematic comparison with results obtained with other models could help shedding light on this part of the residual interaction.

The calculations of NMFP based on Skyrme functionals are particularly fast since the response function is known analytically, thus they could be implemented in astrophysical codes which requires the calculations of NMFP.
In particular, it is found that  the BSk21 functional is able to reproduce the main features of the rather complicated calculations based on CBF, this being a remarkable advantage in terms of computational time.


\section*{Acknowledgments}

This work has been supported by  NewCompstar COST Action MP1304. The work of M.M. has been supported by the Interuniversity Attraction Poles Programme initiated by the Belgian Science Policy Office (BriX network P7/12).
The work of J.N. has been supported by the grant FIS2011-28617-C02-02, Mineco (Spain). The work of  N.C. and S.G. was supported by FNRS (Belgium). 


\clearpage

\begin{appendix}
\begin{widetext}

\section{Skyrme functional}\label{app:functional}

The Skyrme functional adopted in the present article can be written as the sum of two terms $\mathcal{E}_{Skyrme}=\mathcal{E}_{std}+\mathcal{E}_{extra}$.
The first term, $\mathcal{E}_{std}$, corresponds to the functional derived from the \emph{standard} Skyrme effective interaction as discussed in Refs.~\cite{les07,per04} and it reads

\begin{eqnarray}\label{functional:st}
\mathcal{E}_{std}&=& \int d^{3}r  \sum_{t=0,1}\left\{C^{\rho}_{t}[\rho_{0}]\rho_{t}^{2}+C^{s}_{t}[\rho_{0}]\mathbf{s}_{t}^{2}+C^{\Delta\rho}_{t}\rho_{t}\Delta \rho_{t}  +C^{\Delta s}_{t}\mathbf{s}_{t}\Delta \mathbf{s}_{t}+C^{\tau}_{t}(\rho_{t}\tau_{t}-\mathbf{j}_{t}^{2})\right.\nonumber\\
&+& C^{\nabla J}_{t}(\rho_{t}\nabla \mathbf{J}_{t}+\mathbf{s}_{t}\nabla \times\mathbf{j}_{t})+\left. C^{T}_{t} \left( \mathbf{s}_{t}\mathbf{T}_{t}-\sum_{\mu,\nu=x}^{z}J_{t,\mu\nu}J_{t,\mu\nu}\right) \right\} .
\end{eqnarray}

\noindent The coupling constants are related to the coefficient of the Skyrme pseudo-potential as discussed in Ref.~\cite{per04}.
The second term, $\mathcal{E}_{extra}$, corresponds to the contribution of the two extra density dependent terms~\cite{cha09}, which have been added on top of the \emph{standard} Skyrme interaction. It reads 

\begin{eqnarray}\label{functional}
\mathcal{E}_{extra}&=& \int d^{3}r  \sum_{t=0,1}C_{t}^{\Delta\rho,\beta}\rho(r)^{\beta} \rho_{t}(r)\Delta \rho_{t}(r)+C^{\tau,\beta}_{t}\rho(r)^{\beta}\left[ \rho_{t}(r) \tau_{t}(r) -\mathbf{j}_{t}^{2}(r) \right] +C^{\tau,\gamma}_{t}\rho(r)^{\gamma}\left[ \rho_{t}(r) \tau_{t}(r) -\mathbf{j}_{t}^{2}(r) \right] \nonumber\\
&+&C^{\nabla \rho,\beta}_{t}\rho(r)^{\beta}\nabla^{2}\rho_{t}(r)+C^{\nabla \rho,\gamma}_{t}\rho(r)^{\gamma}\nabla^{2}\rho_{t}(r) +C^{\Delta s,\beta}_{t}\rho(r)^{\beta}s_{t}(r)\Delta s_{t}(r)+C^{T,\beta}_{t}\rho(r)^{\beta}\left[s_{t}(r) T_{t}(r) -J^{2}_{t}(r)\right] \nonumber\\
&+&C^{T,\gamma}_{t}\rho(r)^{\gamma}\left[ s_{t}(r) T_{t}(r) -J^{2}_{t}(r)\right]+C^{\nabla s,\beta}_{t} \rho(r)^{\beta}  \nabla^{2} s_{t}(r) +C^{\nabla s,\gamma}_{t} \rho(r)^{\gamma}  \nabla^{2} s_{t}(r). \nonumber\\
\end{eqnarray}

\noindent Note that we have introduced an unusual gradient term $\nabla^{2}\rho_{t}$ to take advantage in the calculations of the residual interaction. Such a term can be eventually re-written as $\rho\Delta\rho$ by a simple integration by part.
The coupling constants of the extra part of the functional $\mathcal{E}_{extra}$ are related to the Skyrme coefficients as

\begin{eqnarray*}
C^{\Delta \rho,\beta}_{0}\rho(r)^{\beta}&=& -\frac{3t_{4}}{32}\rho(r)^{\beta}\\
C^{\Delta \rho,\beta}_{1}\rho(r)^{\beta}&=&\frac{t_{4}}{16}  \left(\frac{1}{2}+x_{4} \right)\rho(r)^{\beta}\\
C^{\tau,\beta}_{0}\rho(r)^{\beta}+C^{\tau,\gamma}_{0}\rho(r)^{\gamma}&=& 3\frac{t_{4}}{16} \rho(r)^{\beta}+\frac{t_{5}}{4}\rho(r)^{\gamma}\left[ \frac{5}{4}+x_{5} \right]\\
C^{\tau,\beta}_{1}\rho(r)^{\beta}+C^{\tau,\gamma}_{1}\rho(r)^{\gamma}&=&- \frac{t_{4}}{8}  \left(\frac{1}{2}+x_{4} \right)\rho(r)^{\beta}+\frac{t_{5}}{8} \rho(r)^{\gamma} \left[\frac{1}{2}+x_{5} \right]\\
C^{\nabla \rho,\beta}_{0}\rho(r)^{\beta}+C^{\nabla \rho,\gamma}_{0}\rho(r)^{\gamma}&=&3\frac{t_{4}}{64}  \rho(r)^{\beta}-\frac{t_{5}}{16} \rho(r)^{\gamma}\left[ \frac{5}{4}+x_{5} \right]\\
C^{\nabla \rho,\beta}_{1}\rho(r)^{\beta}+C^{\nabla \rho,\gamma}_{1}\rho(r)^{\gamma}&=&- \frac{t_{4}}{32} \left(\frac{1}{2}+x_{4} \right)\rho(r)^{\beta}-\frac{t_{5}}{32} \rho(r)^{\gamma} \left[\frac{1}{2}+x_{5} \right]
\end{eqnarray*}

\begin{eqnarray*}
C^{\Delta s,\beta }_{0}\rho(r)^{\beta}&=&\frac{t_{4}}{16}\left( \frac{1}{2}- x_{4}\right) \rho(r)^{\beta}\\
C^{\Delta s,\beta }_{1}\rho(r)^{\beta}&=&\frac{t_{4}}{32}\rho(r)^{\beta}\\
C^{T,\beta}_{0}\rho(r)^{\beta}+C^{T,\gamma}_{0}\rho(r)^{\gamma}&=&\frac{t_{4}}{8}\left( x_{4}-\frac{1}{2}\right) \rho(r)^{\beta}+ \frac{t_{5}}{8}\left( x_{5}+\frac{1}{2}\right)  \rho(r)^{\gamma}\\
C^{T,\beta}_{1}\rho(r)^{\beta}+C^{T,\gamma}_{1}\rho(r)^{\gamma}&=& -\frac{t_{4}}{16}  \rho(r)^{\beta}+ \frac{t_{5}}{16}  \rho(r)^{\gamma} \\
C^{\nabla s,\beta}_{0}\rho(r)^{\beta}+C^{\nabla s,\gamma}_{0}\rho(r)^{\gamma}&=&\frac{t_{4}}{32}\left( x_{4}-\frac{1}{2}\right)   \rho(r)^{\beta}-\frac{t_{5}}{32}\left( x_{5}+\frac{1}{2}\right)  \rho(r)^{\gamma}\\
C^{\nabla s,\beta}_{1}\rho(r)^{\beta}+C^{\nabla s,\gamma}_{1}\rho(r)^{\gamma}&=&- \frac{t_{4}}{64}\rho(r)^{\beta}-\frac{t_{5}}{64} \rho(r)^{\gamma}\,.
\end{eqnarray*}

\section{Coefficients $W_{i}^{(S)}$ }\label{w:pnm}

\noindent Here we give the contribution of the extra term to the coefficients $W^{(S)}_{i=1,2,3}$ for the PNM case. We recall that to obtain the complete expression one should also include the terms related to the second functional derivative of ${\cal E}_{std}$. They are given in Appendix of Ref.~\cite{pas12a}, and are omitted here to have a lighter notation.

\begin{eqnarray}
\frac{1}{2}\bar{W}_{1}^{(0)}&=& {C}^{\tau,\beta} (\beta+1)\beta\rho_{n}^{\beta-1}\tau_{n}+{C}^{\tau,\gamma} (\gamma+1)\gamma\rho_{n}^{\gamma-1}\tau_{n}\nonumber\\
&-&2\mathbf{q}^{2}\left[ {C}^{\Delta\rho,\beta}(\beta+1)\rho_{n}(r)^{\beta}-{C}^{\nabla \rho,\beta}\rho_{n}^{\beta}-{C}^{\nabla \rho,\gamma}\rho_{n}^{\gamma}\right]-\frac{1}{2}\mathbf{q}^{2}\left[ {C}^{\tau,\beta}\rho_{n}^{\beta} +{C}^{\tau,\gamma}\rho_{n}^{\gamma}\right]\,,\\
\frac{1}{2}\bar{W}_{1}^{(1)}&=& -2\mathbf{q}^{2}\left[{C}^{\Delta s,\beta}\rho_{n}^{\beta}-{C}^{\nabla s,\beta}\rho_{n}^{\beta}-{C}^{\nabla s,\gamma}\rho_{n}^{\gamma}  \right]-\frac{1}{2}\mathbf{q}^{2}\left[ {C}^{T,\beta}\rho_{n}^{\beta} +{C}^{T,\gamma}\rho_{n}^{\gamma} \right]\,,
\end{eqnarray}

\begin{eqnarray}
\frac{1}{2}\bar{W}_{2}^{(0)}&=& (1+\beta){C}^{\tau,\beta}\rho_{n}^{\beta} + (1+\gamma){C}^{\tau,\gamma}\rho_{n}^{\gamma}\,,\\
\frac{1}{2}\bar{W}_{2}^{(1)}&=& {C}^{T,\beta}\rho_{n}^{\beta} + {C}^{T,\gamma}\rho_{n}^{\gamma}\,,
\end{eqnarray}

\begin{eqnarray}
\frac{1}{2}\bar{W}_{3}^{(0)}&=&{C}^{\tau,\beta}\beta\rho_{n}^{\beta}+{C}^{\tau,\gamma}\gamma\rho_{n}^{\gamma}\,,\\
\frac{1}{2}\bar{W}_{3}^{(1)}&=& 0\,,
\end{eqnarray}

\noindent where we define the coupling constant ${C}^{X}=C^{X}_{0}+C^{X}_{1}$ and $X=\Delta s, \nabla s,\tau,...$ as done in Ref.~\cite{pas12a}.

\section{Response function $\chi^{(S,M)}_{\text{RPA}}(\mathbf{q},\omega)$ for PNM}\label{chi:pnm}

We write here the complete expressions for the response function in the different spin channels. 
\begin{itemize}

\item For $S=0$ we have
\begin{eqnarray}
\frac{\chi_{HF}}{\chi_{RPA}^{(0)}}&=&1-{\widehat{W}}_{1}^{(0)}\chi_{0}+\frac{1}{2}q^{2}\bar{W}_{3}^{(0)}\chi_{0}+\bar{W}_{2}^{(0)}\left( \frac{q^{2}}{2} \chi_{0} -2k_{F}^{2}\chi_{2}\right)\nonumber\\
&+&[\bar{W}_{2}^{(0)}]^{2}k_{F}^{4}\left[ \chi_{2}^{2}-\chi_{0}\chi_{4}+\left( \frac{m^{*}\omega}{k_{F}^{2}}\right)^{2}\chi_{0}^{2}-\frac{m^{*}}{6\pi^{2}k_{F}}q^{2}\chi_{0}\right]+2\chi_{0}\left( \frac{m^{*}\omega}{q}\right)^{2} \frac{\bar{W}_{2}^{(0)}-\bar{W}_{3}^{(0)}}{1-\frac{m^{*}k_{F}^{3}}{3\pi^{2}}(\bar{W}_{2}^{(0)}-\bar{W}_{3}^{(0)})}\,.
\end{eqnarray}

\item For $S=1,M=\pm1$ we have

\begin{eqnarray}
\frac{\chi_{HF}}{\chi_{RPA}^{(1,\pm1)}}&=&1-{\widehat{W}}_{1}^{(1,\pm1)}\chi_{0}+\frac{1}{2}q^{2}\bar{{W}}_{3}^{(1)}\chi_{0} +\bar{W}_{2}^{(1)} \left\{ \frac{q^{2}}{2}\chi_{0} -2k_{F}^{2}\chi_{2}\right\}\nonumber\\
&+&\left[\bar{W}_{2}^{(1)} \right]^{2}k_{F}^{4}\left\{\chi_{2}^{2}-\chi_{0}\chi_{4}+\left( \frac{m^{*}\omega}{k_{F}^{2}}\right)^{2} \chi_{0}^{2}-\frac{m^{*}}{6\pi^{2}k_{F}}q^{2}\chi_{0}\right\}+2\chi_{0}\left(\frac{m^{*}\omega}{q} \right)^{2}\frac{ \bar{W}_{2}^{(1)}-\bar{W}_{3}^{(1)}}{1-\frac{m^{*}k_{F}^{3}}{3\pi^{2}}\left[\bar{W}_{2}^{(1)}-\bar{W}_{3}^{(1)} \right]}\,.
\end{eqnarray}

\item For $S=1,M=0$ we have

\begin{eqnarray}
\frac{\chi_{HF}}{\chi_{RPA}^{(1,0)}}&=&1-W_{1}^{(1)}\chi_{0}+\frac{1}{2}q^{2}\bar{W}_{3}^{(1)}\chi_{0}+\bar{W}_{2}^{(1)}\left[\frac{q^{2}}{2}\chi_{0}-2k_{F}^{2}\chi_{2} \right]\nonumber\\
&&+[\bar{W}_{2}^{(1)}]^{2}\left[k_{F}^{4}\chi_{2}^{2}-k_{F}^{4}\chi_{0}\chi_{4}+m^{*2}\omega^{2}\chi_{0}^{2}-\frac{k_{F}^{3}m^{*}q^{2}}{6\pi^{2}}\chi_{0} \right]+2\chi_{0}\left(\frac{m^{*}\omega}{q} \right)^{2}\frac{ \bar{W}_{2}^{(1)}-\bar{W}_{3}^{(1)}}{1-\frac{m^{*}k_{F}^{3}}{3\pi^{2}}\left[\bar{W}_{2}^{(1)}-\bar{W}_{3}^{(1)} \right]}\,.
\end{eqnarray}

\end{itemize}

\noindent where we have also defined

\begin{eqnarray}
\widehat{W}_{1}^{(0)}&=&\bar{W}_{1}^{(0)}+\frac{4 q^{4}[{C}^{\nabla J}]^{2}(\beta_{2}-\beta_{3})}{1+q^{2}(\beta_{2}-\beta_{3})[\bar{W}_{2}^{(1)}-\bar{W}_{3}^{(1)}]}\,,\\
\widehat{W}_{1}^{(1,\pm1)}&=&\bar{W}_{1}^{(1)}+\frac{4q^{4}[{C}^{\nabla J}]^{2}(\beta_{2}-\beta_{3})}{1+ q^{2}(\beta_{2}-\beta_{3})[\bar{W}_{2}^{(0)}-\bar{W}_{3}^{(0)}]}\,.
\end{eqnarray}

\noindent Note that the spin-orbit term does not act in the $S=1,M=0$ channel and thus the $\bar{W}_{1}^{(1)}$ is not modified. The definition of the functions $\beta_{i=2,3}$ and $\chi_{i=0,2,4}$ can be found in Ref.~\cite{dav09}.

\end{widetext}
\end{appendix}


\bibliography{biblio}


\end{document}